\documentclass[a4paper,11pt]{article}
\usepackage{jheppub} 
\pdfoutput=1  
\usepackage[T1]{fontenc} 
\usepackage{xcolor}
\definecolor{addedcolor}{rgb}{0.0, 0.7, 0.7}
\newif\ifdraftmode
\draftmodefalse
\newenvironment{added}{%
  \ifdraftmode
    \color{addedcolor}%
  \fi
  \ignorespaces
}{%
  \ignorespacesafterend
}
\newcommand{\addedinline}[1]{%
  \ifdraftmode
    \textcolor{addedcolor}{#1}%
  \else
    #1%
  \fi
}

\title
{Flux Quantization on M-Strings}
\author[a]{P. Banerjee} 
\author[b,c]{H. Sati}
\author[b]{U. Schreiber}

\affiliation[a]{Department of Physics, Virginia Tech}
\affiliation[b]{ Center for Quantum and Topological Systems
(CQTS), New York University Abu Dhabi}
\affiliation[c]{The Courant Institute for Mathematical Sciences, New York University}

\emailAdd{pinakb24@vt.edu}
\emailAdd{hsati@nyu.edu}
\emailAdd{us13@nyu.edu}

\abstract{
  The electric Gauss law in 11D SuGra is famously non-linear, whence its flux quantization must be in nonabelian cohomology. We have previously shown that the minimal admissible choice is 4-Cohomotopy, which in the presence of magnetized M5-probes takes its relative twistorial form.

  Here we discuss how this situation is further refined in the presence of M-string probes on the M5-worldvolume. Based on the superspace formulation of 11D SuGra, we find the nested Bianchi identities by iterating the superembedding construction for super $p$-branes. The resulting probe brane hierarchy (M1 on magnetized M5 in 11D bulk) turns out to admit flux quantization in a doubly-relative form of twisted Cohomotopy, classified by the factorization of the quaternionic Hopf fibration through the twistor fibration.

  The further equivariant refinement of this cohomology theory reduces on A-type singularities to a form of relative Cohomotopy which geometrically engineers Chern-insulator phases on $\mathrm{M5}\cap \mathrm{A}_n$, with the M-string playing the role of gapped nodal lines.   
}

\usepackage[leftcaption]{sidecap}

\usepackage{enumitem}
\setlist[enumerate,1]{label=\textbf{(\roman*)}}
\setlist{
}

\usepackage{lmodern}
\usepackage{anyfontsize}
    
\usepackage{multirow}
\usepackage{hhline}

\usepackage{tabularray}
\UseTblrLibrary{booktabs}

\usepackage{adjustbox}
\usepackage{tcolorbox}

\newtcolorbox{standout}{
  colback=gray!15,
  boxrule=0pt,
  left=.3cm,
  right=.3cm,
  top=.18cm,
  bottom=.18cm,
  boxsep=0pt
}

\usepackage[safe]{tipa}

\usepackage{mathtools}
\usepackage{amssymb}
\usepackage{amsthm}
\usepackage{xfrac}
\usepackage{stmaryrd} % for \sslash
\usepackage{slashed} % for \slashed
\usepackage{scalerel} % for scaling brackets
\usepackage{stackengine} % for incremental scaling

 \newcommand{\bracket}[3]{%
  \stretchleftright
    {#1}
    {%
      \ensurestackMath{\addstackgap[1pt]{#2}}%
      \vrule width 0pt depth 2pt height 0pt
    }
    {#3}%
} 
\newcommand{\bracketmid}[4]{%
  \stretchleftright{#1}{%
    \ensurestackMath{%
      \addstackgap[2pt]{#2}%
      \,\stretchrel*{|}{\addstackgap[2pt]{#2#3}}\,%
      \addstackgap[2pt]{#3}%
    }%
  }{#4}%
}

\theoremstyle{plain}

\theoremstyle{definition}

\theoremstyle{remark}

\usepackage[
]{hyperref}
\usepackage{xurl} % allow line breaks in URLs

\usepackage{cleveref}
\crefname{equation}{}{}
\crefname{section}{\S}{\S\S}
\crefname{subsection}{\S}{\S\S}
\crefname{subsubsection}{\S}{\S\S}
\crefname{definition}{Def.}{Defs.}
\crefname{theorem}{Thm.}{Thms.}
\crefname{corollary}{Cor.}{Cors.}
\crefname{lemma}{Lem.}{Lems.}
\crefname{proposition}{Prop.}{Props.}
\crefname{remark}{Rem.}{Rems.}
\crefname{notation}{Ntn.}{Ntns.}
\crefname{fact}{Fact}{Fact}
\crefname{example}{Ex.}{Exs.}
\crefname{figure}{Fig.}{Figs.}
\crefname{table}{Tab.}{Tabs.}
\crefname{footnote}{ftn.}{ftns.}
\Crefname{footnote}{Ftn.}{Ftns.}

\usepackage{tikz}
\usetikzlibrary{
  cd,
  calc,
  arrows.meta,
  backgrounds,
  decorations,
  decorations.pathmorphing, 
  decorations.pathreplacing
}

\tikzcdset{
  arrow style=tikz, 
  diagrams={
    trim left=
    {([xshift=5pt]current bounding box.west)},
    trim right=
    {([xshift=-5pt]current bounding box.east)},
    baseline=
    {([yshift=-2pt]current bounding box.center)},
    >={
      Computer Modern Rightarrow[
        length=4pt, width=4pt
      ]
    },
    hook/.style={
      {Hooks[right, length=2pt, width=8pt]}->
    },
    hook'/.style={
      {Hooks[left, length=2pt, width=8pt]}->
    },
    shorten=0pt,
  }
}

\definecolor{darkblue}{rgb}{0.05,0.25,0.65}
\definecolor{darkgreen}{RGB}{20,140,10}
\definecolor{lightgray}{rgb}{0.9,0.9,0.9}
\definecolor{darkorange}{RGB}{200,100,5}
\definecolor{darkyellow}{rgb}{.91,.91,0}
\definecolor{lightolive}{RGB}{225, 220, 185}

\let\originalsslash\sslash
\renewcommand{\sslash}{\mathord{\originalsslash}}

\makeatletter
\newcommand{\cpt}{\mathpalette\cpt@inner\relax}
\newcommand{\cpt@inner}[2]{%
  \scalebox{0.5}[0.9]{$#1\cup$}% Scales the cup relative to the current style
  #1\{\infty\}% Typesets the infinity set in the current style
}
\makeatother

\DeclareRobustCommand{\rchi}{{\mathpalette\irchi\relax}}
\newcommand{\irchi}[2]{\raisebox{\depth}{$#1\chi$}}

\tikzset{
  snake left/.style={
    rounded corners,
    to path={
      let \p1 = (\tikztostart.east),
          \p2 = (\tikztotarget.west),
          \p3 = ($(\p1)!0.5!(\p2)$),
          \n1 = {8pt} 
      in
      (\p1)
      -- (\x1 + \n1, \y1)
      -- (\x1 + \n1, \y3)
      -- (\x2 - \n1, \y3) \tikztonodes
      -- (\x2 - \n1, \y2)
      -- (\p2)
    }
  }
}

\tikzset{
  uphordown/.style={
    rounded corners,
    to path={
      let \p1 = (\tikztostart.north),
          \p2 = (\tikztotarget.north),
          \n1 = {max(\y1,\y2) + 8pt}
      in
      (\p1)
      -- (\x1, \n1)
      -- (\x2, \n1) \tikztonodes 
      -- (\p2)
    }
  }
}

\tikzset{
  downhorup/.style={
    rounded corners,
    to path={
      let \p1 = (\tikztostart.south),
          \p2 = (\tikztotarget.south),
          \n1 = {min(\y1,\y2) - 8pt}
      in
      (\p1)
      -- (\x1, \n1)
      -- (\x2, \n1) \tikztonodes 
      -- (\p2)
    }
  }
}

\tikzset{
  rightvertleft/.style={
    rounded corners,
    to path={
      let \p1 = (\tikztostart.east),
          \p2 = (\tikztotarget.east),
          \n1 = {max(\x1,\x2) + 8pt}
      in
      (\p1)
      -- (\n1, \y1)
      -- (\n1, \y2) \tikztonodes 
      -- (\p2)
    }
  }
}

\tikzset{
  leftvertright/.style={
    rounded corners,
    to path={
      let \p1 = (\tikztostart.west),
          \p2 = (\tikztotarget.west),
          \n1 = {min(\x1,\x2) - 8pt}
      in
      (\p1)
      -- (\n1, \y1)
      -- (\n1, \y2) \tikztonodes 
      -- (\p2)
    }
  }
}

\newcommand{\inlinetikzcd}[1]{\begin{tikzcd}[sep=small, ampersand replacement=\&]#1\end{tikzcd}}

\newcommand{\CyclicGroup}[1]{\mathbb{Z}_{/#1}}

\renewcommand{\setminus}{-}

\newcommand{\defneq}{\equiv}

\newcommand{\StringVielbein}{\varepsilon}
\newcommand{\StringGravitino}{\rchi}
\newcommand{\StringSpinConnection}{\varpi}

\newcommand{\slashH}{\;\slash \hspace{-5.5pt}\tilde H}

\newcommand{\CoefficientMatrix}{M}

\newcommand{\shape}{%
  {\mathord{\scalerel*{\raisebox{0.1ex}{\textesh}}{f}}%
  \mkern 2mu}
}

\begin{document}
%%%%%%%%%%%%%%%%%%%%%%%%%

%%%%%%%%%%%%%%
\maketitle
%%%%%%%%%%%%%%

%%%%%%%%%%%%%%%%%%%%%%%%%%
\section{Introduction}
%%%%%%%%%%%%%%%%%%%%%%%%%%

%%%%%%%%%%%%%%%%
\subsection
{The Problem}
%%%%%%%%%%%%%%%%

Despite the relevance of topological quantum effects in general (\cite{Simon2023}, notably in potential applications to topological quantum hardware \cite{Nayak2008}) and in particular in string/M-theory \cite{Duff1999World,MiemiecSchnakenburg2006}, the global topological completion of (higher) gauge fields coupling to branes (\cite[\S 7]{Fre2013Branes,Lu2025}) has not received sufficient attention.

For example, the famous conjecture that RR-flux is quantized in K-theory (cf. \cite{MooreWitten2000,Freed2000,Grady2022} and \cref{ChargeQuantizationInKTheory} below) tacitly admits that besides magnetic flux also (Hodge dual) electric fluxes are to be charge-quantized. But this insight has been applied to other flux species only recently (review in \cite{SS25-Flux}), where it requires new mathematical methods (such as developed in \cite{FSS23-Char,SS25-TEC}), since electric Gauss laws are generically non-linear, precluding their quantization in Whitehead-generalized abelian cohomology theories like K-theory (we briefly review this below in \cref{OnProperFluxQuantization}).

Yet more pronounced subtleties appear when bulk fluxes are probed by branes that carry their own worldvolume flux densities: For instance, whichever quantization law is chosen for the 4-flux $G_4$ in the bulk $X$ of 11D Supergravity (cf. \cite[\S 3.1.3]{MiemiecSchnakenburg2006}), which is already subtle (\cite{Witten1997flux,DFM2007,FSS20-H}), it must affect the quantization of the (itself rather subtle) self-dual 3-form flux $H_3$ on M5-brane embeddings  $\inlinetikzcd{\Sigma \ar[r, hook, "{ \Phi }"] \& X}$ (cf. \cite[\S 3]{Duff1999World}), due to the \emph{relative Gauss law}, $\mathrm{d}\, H_3 = \Phi^\ast G_4$ (\cite[(36)]{HoweSezgin1997}\cite[(5.75)]{Sorokin2000}), that relates the two; but no real method to address this situation had been considered until recently \cite{GSS25-M5,FSS21-Hopf,FSS21-StrStruc} (but cf. \cite{Kalkkinen_2003,Figueroa_O_Farrill_2001,Stanciu2000}).

Here we explore the iteration of this subtlety, namely the global completion of 11D SuGra by flux quantization in the presence of a \emph{sequence} of probe brane embeddings. We also include an ``M-string'' 
(cf. \cite[\S 11]{DuffLuXin1994}\cite{HaghighatEtAl2015}) probing the M5-brane probe of the 11D bulk (cf. \cref{MStringSchematics}):
\begin{equation}
  \begin{tikzcd}
    \text{M-string}
    \ar[r, hook, "{ \phi }"]
    &
    \text{M5-brane}
    \ar[r, hook, "{ \Phi }"]
    &
    \text{11D bulk}
    \mathrlap{\,.}
  \end{tikzcd}
\end{equation}

Besides showcasing (in \cref{TheFluxQuantizationOfMOnM5In11D}) the algebro-topological technology which allows proper flux quantization of such rich situations, we will demonstrate (in \cref{OnOrbifolding}) that, on A-type orbi-singularities, the global completion of 11D SuGra with such sequential probe branes naturally admits \emph{geometric engineering} (cf. \cite{Duplij2017}) of anyonic topological quantum effects recently seen in experiment (in the conclusion \cref{Conclusion} below).

\begin{added}

%%%%%%%%%%%%%%
\subsection
{\addedinline{The Wider Picture}}
\label
{TheWiderPicture}
%%%%%%%%%%%%%%
 We expand on the general issue addressed here, of global completion of higher gauge theories by electromagnetic flux quantization. Further background exposition is given in \cite{SS26-HigherGauge,SS25-Flux,Schreiber2025}.

  Traditionally, much of fundamental theoretical physics has been and still is tacitly considered only on a single chart of spacetime. For instance, configurations of the C-field (``3-index photon'') in 11D supergravity are routinely identified with differential 3-forms, $C_3$, and configurations of the B-field (``self-dual tensor field'') on the M5-brane are routinely identified with differential 2-forms $B_2$, etc. 
  This habit is tightly connected to the ubiquitous idea that field theories would be defined by Lagrangian densities that are wedge polynomials in these \emph{gauge potential} differential forms and their de Rham differentials.

  That this is not actually the case remains well-appreciated only in the simplest example of the A-field, namely the usual electromagnetic field whose gauge potential is locally a 1-form $A_1$.

\paragraph
{The Case of Electromagnetism.}
  In this case, going back to the seminal article by Dirac 1931 \cite{Dirac1931}, it is well-known now that $A_1$ is only the local data of the electromagnetic field, whose global completion is given by a differential-topological structure equivalently known as a \emph{connection} on a complex line bundle, or a 2-cocycle in \emph{differential integral cohomology}.

  A modern way to say what this classical statement means is:
  In addition to the electromagnetic flux density $F_2$ (locally the differential of $A_1$), the electromagnetic field, globally on spacetime $X^{1,d}$, involves a classifying map $ \inlinetikzcd{X^{1,d} \ar[r, "{ \rchi }"] \& B^2 \mathbb{Z}}$ to the \emph{Eilenberg-MacLane space} $B^2 \mathbb{Z}$ (often denoted $K(\mathbb{Z},2)$, classifying integral 2-cohomology), and the collection of local gauge potentials $A_1$ are what exhibits an identification of the de Rham 2-cocycle $F_2$ with the integral 2-cocycle $\rchi$ in real cohomology (cf. \cite[\S 9, Prop. 9.5]{FSS23-Char}):
\begin{equation}
\label
{OrdinaryFluxQuantization}
  \begin{tikzcd}[
    sep=30pt
  ]
    \ast
    \ar[
      r,
      dashed,
      "{ 
        \rchi 
      }"{swap, pos=.7,name=charge},
      "{
        \text{\color{darkblue}charge}
      }"{pos=.7}
    ]
    \ar[
      d,
      dashed,
      "{
        F_2
      }"{pos=.7,name=flux},
      "{
        \text{\color{darkblue}flux}
      }"{swap, pos=.7}
    ]
    & 
    \mathrm{Map}\bracket({
      X^{1,d},
      B^2 \mathbb{Z}
    })
    \ar[
      d,
      "{
        \mathbf{ch}
      }"
    ]
    \ar[
      from=charge,
      to=flux,
      Rightarrow,
      dashed,
      "{
        \text{\color{darkblue}potentials}
      }"{sloped},
      "{ \widehat{A_1} }"
    ]
    \\[15pt]
    \mathbf{\Omega}^2_{\mathrm{cl}}
    \bracket({
      X^{1,d}
    })
    \ar[
      r,
      "{
        \eta^{\shape}
      }"
    ]
    &
    \mathrm{Map}\bracket({
      X^{1,d},
      B^2 \mathbb{R}
    })
    \mathrlap{\,.}
  \end{tikzcd}
\end{equation}
Here the hat on $\widehat{A_1}$ indicates that this differential 1-form is only locally defined, namely on an open cover $\widehat{X}^{1,d}$ of spacetime.

  As indicated in \cref{OrdinaryFluxQuantization}, the class $[\rchi] \in H^2\bracket({X^{1,d}; \mathbb{Z}})$ here is the total \emph{charge} which sources the electromagnetic field, carried by the ``0-branes'' of electromagnetism, the \emph{magnetic monopoles}. The diagram implies that this is integrally ``quantized'' in that the non-torsion part of this integral charge is reflected in the flux density $F_2$ being an \emph{integral form}, in that its  total flux through any sphere is an integer.

  This global completion of the Maxwell field has crucial physical impact: It describes not just the  effect of (still hypothetical) magnetic monopoles, but also the experimentally observed flux quanta through Abrikosov vortices (the ``solitonic 1-branes'' of electromagnetism, cf. \cite[\S 2.1]{SS25-Flux}) in type II superconductors, and, last not least, the effective gauge field of anyons in fractional quantum Hall systems, in the guise of $\mathrm{U}(1)$-Chern-Simons fields (cf. \cite{GuadagniniThuillier2008}). In the latter case, the flux density vanishes on-shell, $F_2 = 0$, and the field content is hence \emph{all} in the global completion. Below in \cref{OnProperFluxQuantization,OnMStringFluxQuantization} we are interested in seeing this kind of phenomenon realized on M5-branes and M-strings. 

  But the point of general  importance here is that this global completion of the electromagnetic field is all controlled by the choice of the \emph{classifying space} $B^2 \mathbb{Z}$, encoding traditional \emph{Dirac charge quantization}.

  In fact, one could choose other classifying spaces $\mathcal{A}$ (cf. \cite[Ex. 3.9]{SS25-Flux}), which induce other global completions of the electromagnetic field by variant flux quantization laws (and hence imply potentially other experimental predictions): What makes the globalization construction \cref{OrdinaryFluxQuantization} work is that:
\begin{enumerate}
  \item The Bianchi identity $\mathrm{d}F_2 = 0$ induces a single real 2-cohomology class, hence classified by $B^2\mathbb{R}$, so that we have the bottom map $\eta^{\shape}$.
  \item 
  The ``realification'' ($\mathbb{R}$-rationalization, cf. \cite[\S 5]{FSS23-Char}) of $\mathcal{A}$ coincides with $B^2 \mathbb{R}$, so that we have the \emph{character map} $\mathbf{ch}$ on the right.
\end{enumerate}

  A key demand on the ``right'' choice of a flux quantization law $\mathcal{A}$ is global consistency of probe branes charged under these background fluxes: For instance Dirac's flux quantization \cref{OrdinaryFluxQuantization} was motivated and is justified by the fact (cf. \cite{Alvarez1985}\cite[\S 2]{Gawedzki1988}) that it gives consistent global meaning to the topological term 
  \begin{equation}
  \label
  {ChargedParticleTopologicalTerms}
    \text{``{}}
    \exp\bracket({
    2\pi
    \mathrm{i} \, \int_{\Sigma^{1,0}} \phi^\ast {A_1}
    })
    \text{{}''}
  \end{equation}
  in the exponentiated action functional for an electron worldline $\inlinetikzcd{\Sigma^{1,0} \ar[r, "{ \phi }"] \& X^{1,d}}$,  by understanding it as the \emph{holonomy} functional defined by the differential cocycle \cref{OrdinaryFluxQuantization}.

  In the same vein one should ask, for instance: \emph{What flux-quantization condition on the combined C-field \textup{(locally: $C_3$, $C_6$)} in 11D SuGra and the tensor field \textup{(locally: $B_2$)} on an M5-brane} worldvolume
  $\inlinetikzcd{ \Sigma^{1,5} \ar[r, "{ \Phi }"] \& X^{1,10} }$ 
  ensures that the M5's exponentiated Hopf-WZW action
  \cite[(1)]{BandosEtAl1997}
  \begin{equation}
  \label
  {HopfWZTerm}
    \text{``{}}
    \exp\bracket({ 
    4\pi \mathrm{i}
    \,
    \int_{\Sigma^{1,5}} \bracket({ \Phi^\ast C_6 - \tfrac{1}{2} \mathrm{d}B_2 \wedge \Phi^\ast C_3 }) 
    })
    \text{{}''}
    \mathrlap{\,,}
  \end{equation} 
  \emph{is globally well-defined?} 

  This question was only answered in \cite{FSS21-Hopf}, using the method \cite{FSS23-Char,SS25-Flux} that we are concerned with here and briefly survey now.

%%%%%%%%%%%%
\paragraph
{The Case of Higher Gauge Fields.}
%%%%%%%%%%%%
In fairly straightforward generalization of the above situation of electromagnetism, we may now ask for global completion of \emph{any} higher gauge field of Maxwell type, including those with non-linear Bianchi identities, where the flux densities are not necessarily closed but have differentials that are wedge-polynomials in the other flux densities.

  Briefly, for the higher gauge field sectors of 11D supergravity and its descendants:
  \begin{enumerate}
  \item
  there is an $L_\infty$-algebra $\mathfrak{a}$ such that the duality-symmetric Bianchi identities equivalently characterize closed (i.e., flat, Maurer-Cartan) $\mathfrak{a}$-valued differential forms in $\Omega^1_{\mathrm{cl}}\bracket({X^{1,d}; \mathfrak{a} })$ (cf. \cite[\S 2.5, 3.1]{SS25-Flux}),

  \item
  any choice of classifying space $\mathcal{A}$ whose $\mathbb{R}$-rationalization $L^{\mathbb{R}}\mathcal{A}$ has Whitehead brackets $\mathfrak{l}\mathcal{A} \simeq \mathfrak{a}$ provides an admissible flux-quantization (cf. \cite[\S 3.2]{SS25-Flux}),

  \item 
  in that there exists solid maps as follows, in generalization of \cref{OrdinaryFluxQuantization}, whence the corresponding globally completed gauge field configurations are given by a ``cone'' of dashed maps (cf. \cite[\S 3.3]{SS25-Flux}):
\begin{equation}
\label
{GeneralFluxQuantization}
  \begin{tikzcd}[
    sep=30pt
  ]
    \ast
    \ar[
      r,
      dashed,
      "{ 
        \rchi 
      }"{swap, pos=.7,name=charge},
      "{
        \text{\color{darkblue}charge}
      }"{pos=.7}
    ]
    \ar[
      d,
      dashed,
      "{
        \vec F
      }"{pos=.7,name=flux},
      "{
        \text{\color{darkblue}flux}
      }"{swap, pos=.7}
    ]
    & 
    \mathrm{Map}\bracket({
      X^{1,d},
      \mathcal{A}
    })
    \ar[
      d,
      "{
        \mathbf{ch}
      }"
    ]
    \ar[
      from=charge,
      to=flux,
      Rightarrow,
      dashed,
      "{
        \text{\color{darkblue}potentials}
      }"{sloped},
      "{ \widehat{A} }"
    ]
    \\[15pt]
    \mathbf{\Omega}^1_{\mathrm{cl}}
    \bracket({
      X^{1,d};
      \mathfrak{a}
    })
    \ar[
      r,
      "{
        \eta^{\shape}
      }"
    ]
    &
    \mathrm{Map}\bracket({
      X^{1,d},
      L^{\mathbb{R}}
      \mathcal{A}
    })
    \mathrlap{\,.}
  \end{tikzcd}
\end{equation}
\end{enumerate}
For the present purpose we do not need the full details of where such diagrams live (namely in cohesive homotopy theory) and how exactly the objects and maps are defined (namely via the fundamental theorem of dg-algebraic rational homotopy theory).  The key point here is just that:

\begin{standout}
Any choice of $\mathcal{A}$ which is admissible, $\mathfrak{l}\mathcal{A} \simeq \mathfrak{a}$, gives a global completion of the higher gauge theory, in which the topological (brane) charges correspond to homotopy classes of maps $\inlinetikzcd{X^{1,d} \ar[r, dashed] \& \mathcal{A}}$.
\end{standout}
If one imagines that the physics to be described is fixed (e.g.: the ``M-theory'' completion of 11D SuGra) then any such choice is a \emph{hypothesis} about the correct global (infrared) description.

In \cref{OnProperFluxQuantization} we go through examples of how admissible such classifying spaces $\mathcal{A}$ (and more generally: classifying fibrations) are found.

%%%%%%%%%%%%%
\subsection
{\addedinline{From M-Theory to Math}}
%%%%%%%%%%%%%

Concretely, the duality-symmetric flux Bianchi identities of 11D SuGra are, famously as shown on the left here (cf. \cite[\S 3.1.3]{MiemiecSchnakenburg2006}\cite[Thm. 3.1]{GSS24-SuGra}):
\begin{equation}
\label
{ClosedLS4ValuedForms}
  \left\{
  \begin{aligned}
    & G_4
    \\
    & G_7
  \end{aligned}
  \;\middle\vert\;
  \begin{aligned}
    \mathrm{d}\,
    G_4 & = 0
    \\
    \mathrm{d}\, 
    G_7 & =
    \tfrac{1}{2}
    G_4 \wedge G_4
    \mathrlap{\,.}
  \end{aligned}
  \right\}
  \simeq
  \Omega^1_{\mathrm{cl}}\bracket({
    X^{1,d};
    \mathfrak{l}
    S^4
  })
  \mathrlap{\,.}
\end{equation}
As indicated on the right,
a standard fact (cf. \cite[\S 1.2]{Menichi2015}\cite[Ex. 5.3]{FSS23-Char}) in rational homotopy theory implies (cf. \cite[p. 21]{SS25-Flux}) that pairs $(G_4, G_7)$ satisfying these identities are equivalently closed differential forms with coefficients in the Whitehead-bracket $L_\infty$-algebra $\mathfrak{l}S^4$ of the 4-sphere. 

This $L_\infty$-algebra is also known as the \emph{M-theory gauge algebra} \cite[(2.6)]{CremmerEtAl1998}\cite[(3.4)]{LavrinenkoEtAl999}\cite[(75)]{KalkkinenStelle2003}\cite[(86)]{BandosEtAl2004}\cite[(4.9)]{Sati2010}, which is the free graded Lie algebra $\mathbb{R}_{_L}\langle-\rangle$ on generators $v_3$ and $v_6$ subject to one nontrivial Lie bracket relation:
\begin{equation}
  \mathfrak{l}
  S^4
  \,\simeq\,
  \mathbb{R}_{_{L}}\langle
    v_3, v_6
  \rangle
  \big/
  \Big(
    [v_3, v_3] = v_6
  \Big)
  \mathrlap{\,.}
\end{equation}

It follows with \cref{GeneralFluxQuantization} that the set of possible global (infrared) completions of the supergravity C-field is: 
\begin{equation}
\label
{SetOf11DSuGraCompletions}
  \left\{
  \substack{
    \text{infrared completions}
    \\
    \text{of 11D supergravity}
  }
  \right\}
  \simeq
  \Big\{
  \text{homotopy types $\mathcal{A}$}
  \,{\Big\vert}\,
  \mathfrak{l}\mathcal{A} \simeq \mathfrak{l}S^4
  \Big\}
  \mathrlap{\,.}
\end{equation}

Of the infinitude of distinct possible choices, we mention three:
\begin{enumerate}

\item
the homotopy fiber space
of the rationalized cup square on $B^4 \mathbb{Z}$:
\begin{equation}
\label
{TheBaselineChoice}
  \mathcal{A}
  \defneq
  \mathrm{hfib}\Big(
    \inlinetikzcd{
      B^4 \mathbb{Z}
      \ar[
        rr,
        "{
          (-)^{\cup_2}  
        }"
      ]
      \&\&
      B^8 \mathbb{Z}
      \ar[r]
      \&
      B^8 \mathbb{Q}
    }
  \Big)
  \mathrlap{\,.}
\end{equation}

This choice  completes the $G_4$-flux to a differential integral 4-cocycle (a ``$\mathrm{U}(1)$-bundle 2-gerbe connection'', as direct higher generalization of how $F_2$ is quantized in \cref{OrdinaryFluxQuantization} by a $\mathrm{U}(1)$-bundle connection)  and essentially imposes the non-linear second Bianchi identity \cref{ClosedLS4ValuedForms} only on $G_7$ itself, not though on its integral charge structure.

Essentially this choice is tacitly the C-field flux-quantization traditionally considered in the literature  
\cite{AschieriJurco2004,DFM2007,FSS15-ModuliStack}.

Of course, with $G_7$ remaining essentially unquantized with this choice, it cannot imply the (Page) charge quantization of M2-branes.
In any case, there are other choices one may consider:

\item
The 4-sphere itself:
\begin{equation}
\label
{ChoiceOfS4}
  \mathcal{A}
  \defneq
  S^4
  \mathrlap{\,.}
\end{equation}
This choice, originally proposed in \cite[\S 2.5]{Sati2018}, is the \emph{minimal} choice in number of cells (hence it is ``universal'' in that its charges map into the charges for all other choices). 

In \cite{FSS20-H,FSS21-Hopf,SS23-Mf} it was shown that this choice implies a whole list of subtle topological conditions that are expected in M-theory, notably it implies:
\begin{enumerate}
\item
\cite[Prop. 3.13]{FSS20-H}: the $\tfrac{1}{4}p_1$-shifted integral quantization of $G_4$ argued in \cite{Witten1997flux},

\item
\cite[Thm. 4.8]{FSS21-Hopf}: global consistency of the  M5's topological term \cref{HopfWZTerm}.

\end{enumerate}

Closely related to the second point is the more immediate statement that $\mathcal{A} \defneq S^4$ quantizes not just the charge of black M5-branes but also of black M2-branes, because,
\begin{equation}
  \begin{aligned}
  \pi_0\, \mathrm{Map}\bracket({
    \mathbb{R}^{1,10}
    \setminus
    \mathbb{R}^{1,5},
    S^4
  })
  & 
  \simeq
  \pi_4\bracket({S^4})
  \simeq
  \mathbb{Z}
  \,,
  \\
  \pi_0\, \mathrm{Map}\bracket({
    \mathbb{R}^{1,10}
    \setminus
    \mathbb{R}^{1,2},
    S^4
  })
  & 
  \simeq
  \pi_7\bracket({S^4})
  \simeq
  \mathbb{Z}
  \mathcolor{gray}{
    \oplus \mathbb{Z}_{/12}
  }
  \mathrlap{\,.}
  \end{aligned}
\end{equation}

These results support the hypothesis that \cref{ChoiceOfS4} is the ``right'' choice of global completion for the purpose of ``M-theory'' and as such the choice \cref{ChoiceOfS4} has been called \emph{Hypothesis H} (\cite{FSS20-H}, since it implies that M-brane charges are in the nonabelian cohomology called \emph{co-Homotopy} \cite[\S VII]{STHu59}\cite[Ex. 2.7]{FSS23-Char}, hence ``H-cohomology'' analogous to the more familiar K-theoretic ``K-cohomology''). 

This is the choice we will be concerned with below. But there are still other choices of interest:

\item
The homotopy fiber of the squared second $\mathrm{SU}(2)$-Chern class:
\begin{equation}
  \mathcal{A}
  \defneq
  \mathrm{hfib}\Big(
    \inlinetikzcd{
      B \mathrm{SU}(2)
      \ar[
        r,
        "{ c_2 }"
      ]
      \&
      B^4 \mathbb{Z}
      \ar[
        rr,
        "{
          (-)^{\cup_2}
        }"
      ]
      \&\&
      B^8 \mathbb{Z}
    }
  \Big)
  \mathrlap{\,.}
\end{equation}
Since $S^4 \simeq \mathbb{H}P^1$ and $B \mathrm{SU}(2) \simeq \mathbb{H}P^\infty$ (quaternionic projective spaces), one may understand this choice as a kind of hybrid of the previous two choices \cref{ChoiceOfS4,TheBaselineChoice}.

Its noteworthy property is \cite{BaSS26-UnstableK} that under dimensional reduction to 10D, it implies a flux-quantization of type IIA supergravity which is a form of twisted unstable K-theory. 

The topological charges predicted by this choice are actually indistinguishable from those predicted by Hypothesis H on spacetime compactifications of the form $X^{1,10} \simeq \mathbb{R}^{1,3} \times X^7$, since the canonical comparison map
\begin{equation}
  \begin{tikzcd}[
    column sep=-3pt
  ]
    S^4 
    \ar[r]
    &[25pt]
    \mathrm{hfib}
    \Big(
      B \mathrm{SU}(2)
      \ar[
        r,
        "{ c_2 }"
      ]
      &[15pt]
      B^4 \mathbb{Z}
      \ar[
        r,
        "{ (-)^{\cup_2} }"
      ]
      &[20pt]
      B^8 \mathbb{Z}
    \Big)
  \end{tikzcd}
\end{equation}
is a 7-equivalence \cite[Thm. 2.6]{BaSS26-UnstableK}.
\end{enumerate}

This baseline understanding of global completion of 11D supergravity by C-field flux quantization generalizes to various kinds of further structure, notably to the presence of probe M5-branes \cite{GSS25-M5,FSS21-StrStruc} and to orbifold spacetimes \cite{SS20-Tad,SS25-Seifert}.

\paragraph
{Outline.}
Our aim here is to discuss the further refinement of flux-quantization to the situation where M5-probes of A-type singularities are themselves probed by M-strings. To this end, we now:
\begin{description}
\item[(\cref{OnMStringSuperembedding})]
derive the relevant Bianchi identities in the presence of M-string probes,

\item[(\cref{OnProperFluxQuantization})] recall more technical detail of flux-quantization with super $p$-brane probes,

\item[(\cref{OnMStringFluxQuantization})]
construct an admissible flux quantization for M-strings probing M5-branes probing an 11D bulk with an A-type orbi-singularity.
\end{description}

\end{added}

%%%%%%%%%%%%%%%%%%%%%%%%%%%%%%%%%%%%%
\section{M-String Superembedding into M5}
\label
{OnMStringSuperembedding}
%%%%%%%%%%%%%%%%%%%%%%%%%%%%%%%%%%%%%

To start with, we apply the \emph{super-embedding formalism} 
(going back to \cite{HoweSezginWest1998,HoweRaetzelSezgin1998,Sorokin2000})
to the construction of M-string probes (cf.  \cite{HaghighatEtAl2015} going back to \cite[\S 11]{DuffLuXin1994}) of M5-brane probes of the 11D SuGra bulk. We follow the super-embedding construction of the M5-brane itself, as laid out in \cite{GSS25-M5,GSS25-Embedding}, following \cite{HoweSezgin1997}\cite[\S 5.2]{Sorokin2000}.

\begin{added}
\paragraph
{Superembedding Formalism.}
For context, we briefly review the idea of the super-embedding construction of the M5-brane itself.

Historically (cf. \cite{Sorokin2001}), the superstring was famously first found in its Ramond-Neveu-Schwarz (RNS) formulation. While this has manifest supersymmetry on the worldsheet, the supersymmetry of its spacetime spectrum is not manifest and came as a surprise (that fueled the early interest in string theory). Later, the Green-Schwarz (GS) superstring model made spacetime supersymmetry manifest, but now at the cost of losing manifest worldsheet supersymmetry and introducing a somewhat more unwieldy ``$\kappa$-symmetry'' constraint.
The \emph{superembedding} formulation arose as a  synthesis of the lessons learned from the RNS and the GS models, originally called the \emph{doubly supersymmetric geometrical approach} \cite{BandosEtAl1995}: 

Here the string worldsheet is modeled as a supermanifold, of super-dimension $(1,1\vert 8 \cdot \mathbf{2})$, as is the target spacetime (for type IIA, say), of super-dimension $(1,9\vert \mathbf{16} \oplus \overline{\mathbf{16}})$, and the worldvolume fields are understood as constituting supergeometric embeddings (immersions, really) of supermanifolds:
\begin{equation}
  \inlinetikzcd{
    \Sigma^{1,1\vert 8 \cdot \mathbf{2}}
    \ar[
      rr,
      "{ \Phi }"
    ]
    \&\&
    X^{1,9\vert \mathbf{16} \oplus \overline{\mathbf{16}}}
    \mathrlap{\,.}
  }
\end{equation}
Remarkably, the equations of motion of the superstring then turn out to be equivalent to the requirement that this super-embedding is ``tangent space-wise $\sfrac{1}{2}$-BPS'' in a suitable sense (stated in a moment, originally known as the \emph{geometrodynamical condition}, then as the \emph{basic embedding condition} and eventually as the \emph{superembedding condition} \cite{HoweSezginWest1998,Sorokin2000}, overview in \cite[Rem. 2.24]{GSS25-M5}). With this, supersymmetry of the model is manifest both on the worldsheet and on the target space, and the GS ``$\kappa$-symmetry'' is geometrized as the evident super-diffeomorphism invariance of the formulation \cite[\S4.3]{Sorokin2000}.

Analogous statements may be made for essentially all the super $p$-branes one encounters in string/M-theory. This is particularly striking for those branes that carry higher flux densities (``tensor fields'') on their worldvolume, such as the M5-brane with its $H_3$ flux: Its superembeddings are of the form
\begin{equation}
  \inlinetikzcd{
    \Sigma^{1,5\vert 2\cdot \mathbf{8}}
    \ar[
      rr,
      "{ \Phi }"
    ]
    \&\&
    X^{1,10\vert \mathbf{32}}
    \mathrlap{\,,}
  }
\end{equation}
and a remarkably subtle analysis of the superembedding condition, combined with the supertorsion constraint of the 11D SuGra target, shows (\cite{HoweSezgin1997}, cf. \cite[Prop. 3.18]{GSS25-M5}) that the naive form of the Bianchi identity,
\begin{equation}
  \mathrm{d}
  \,
  H^s_3
  =
  \Phi^\ast
  G^s_4
\end{equation}
but imposed on suitable supergeometric incarnations $(-)^s$ of the flux densities, is equivalent to the equations of motion on $H_3$, including its notorious non-linear self-duality property (cf. \cite[Rem. 3.19]{GSS25-M5}). Nontrivial explicit solutions of M5-superembeddings are exhibited in \cite{GSS25-Embedding}.

Such an equivalence between (i) duality-symmetric Bianchi identities on super-flux densities over superspace and (ii) super-gravity/brane equations of motion also holds for 11D supergravity itself (\cite[Thm. 3.1]{GSS24-SuGra}). This is noteworthy in view of the fact that these Bianchi identities are also the input datum for the flux quantization procedure (as indicated in \cref{TheWiderPicture} and further discussed in \cref{OnSuperFluxQuantization} below).

Therefore, we now begin with indicating the analogous analysis for the superembedding of M-strings \emph{into} M5-branes.

%%%%%%%%%%%%
\paragraph
{The Superembedding condition}
%%%%%%%%%%%%

More in detail, the strong form of the superembedding condition that is relevant here (cf. \cite[Rem. 3.20]{GSS25-M5}) says that the superembedding must be \emph{$\sfrac{1}{2}$-BPS} in the following sense \cite[Def. 2.19, Lem. 2.20]{GSS25-M5} (for relevant background on supergeometry see \cite{Varadarajan2004}\cite[\S 2]{GSS24-SuGra}): 

On  the super-vector space $\mathbb{R}^{1,d\vert \mathbf{N}}$ which is the local model of superspacetime $X^{1,d\vert \mathbf{N}}$, let 
\begin{equation}
\label
{ProjectionOperator}
  P
  :
  \inlinetikzcd{
    \mathbb{R}^{1,d\vert \mathbf{N}}
    \ar[
      r,
      ->>
    ]
    \&
    \mathbb{R}^{1,p\vert \mathbf{n}}
    \ar[
      r,
      hook
    ]
    \&
    \mathbb{R}^{1,d\vert \mathbf{N}}
  }
\end{equation}
be the super-projection operator onto the local model space $\mathbb{R}^{1,p\vert \mathbf{n}}$ of the super $p$-brane, with $n = N/2$ (the \emph{tangential} projection on the fixed space of a \emph{$p$-brane involution} \cite[Def. 4.4]{HSS19}, exhibiting the local halving of the ``number of supersymmetries''). Write $P_{\!\!{}_\perp\!}$ for the orthogonal super-projector (the \emph{transversal} projection).

Then a super-immersion
\begin{equation}
  \inlinetikzcd{
    \Sigma^{1,p\vert \mathbf{n}}
    \ar[
      rr,
      "{ \Phi }"
    ]
    \&\&
    X^{1,d\vert \mathbf{N}}
  }
\end{equation}
is said to satisfy the \emph{$\sfrac{1}{2}$-BPS embedding condition} \cite[Def. 2.19, Lem. 2.20]{GSS25-M5} iff there exists an orthonormal super-coframe $\bracket({ \bracket({E^a})_{a=0}^d, \bracket({ \Psi^\alpha })_{\alpha_1}^{N} })$ on $X^{1,d\vert \mathbf{N}}$ 
and a \emph{super-shear map}
\begin{equation}
\label
{TheShearMap}
  \mathrm{Sh}
  \in
  C^\infty\bracket({
    \widetilde{\Sigma};
    \mathrm{Hom}_{\mathbb{R}}\bracket({
      P\bracket({
        \mathbb{R}^{1,d\vert \mathbf{N}}
      }),
      P_{\!\!{}_\perp\!}\bracket({
        \mathbb{R}^{1,d\vert \mathbf{N}}
      }),      
    })
  })
\end{equation}
(defined on the open cover $\widetilde{\Sigma}$ of the worldvolume on which its super-coframe is defined, cf. \cite[\S 2.1]{GSS25-M5})
such that:
\begin{enumerate}
  \item
  The pullback of the tangential part of the target super-coframe is a super-coframe $(e,\psi)$ on the brane:
  \begin{equation}
  \label
  {TangentialSuperembeddingCondition}
    \Phi^\ast\bracket({
      P\bracket({
        \bracket({E^a})_{a=0}^d,
        \bracket({\Psi^\alpha})_{\alpha=1}^N
      })
    })
    =
    \bracket({
      \bracket({e^a})_{a=0}^{p},
      \bracket({ \psi^\alpha })_{\alpha=1}^{n}
    }).
  \end{equation}
  
  \item
  The pullback of the transversal part is expanded in this coframe by the given shear map \cref{TheShearMap}:
  \begin{equation}
  \label
  {TransversalSuperembeddingCondition}
    \begin{aligned}
    \Phi^\ast
    \bracket({
      P_{\!\!{}_\perp\!}\bracket({
        \bracket({E^a})_{a=0}^d,
        \bracket({\Psi^\alpha})_{\alpha=1}^N
      })
    })
    & =
    \mathrm{Sh} \cdot 
    \Phi^\ast
    \bracket({
      P\bracket({
        \bracket({E^a})_{a=0}^d,
        \bracket({\Psi^\alpha})_{\alpha=1}^N
      })
    })
    \\
    & 
    \underset{\mathclap{
      \text{\cref{TangentialSuperembeddingCondition}}
    }}{=}
    \mathrm{Sh} \cdot 
    \bracket({
        \bracket({e^a})_{a=0}^p,
        \bracket({\psi^\alpha})_{\alpha=1}^n
    })
    \end{aligned}
  \end{equation}
  and \emph{equivariantly so} with the transversal $\mathrm{Spin}(d-p)$-action.
\end{enumerate}

Remarkably, the transverse fermionic shear in \cref{TransversalSuperembeddingCondition} is, when it exists, the origin of nonvanishing worldvolume flux densities, in addition to the ``embedding fields'' that constitute the tangential component \cref{TangentialSuperembeddingCondition}. But the transversal equivariance condition on the shear means, by Schur's lemma, that it can only be non-vanishing if the $\mathrm{Spin}(d-p)$-representations $P(\mathbf{N})$ and $P_{\!\!{}_\perp\!}(\mathbf{N})$ share isomorphic irrep summands (cf. \cite[Rem. 2.2]{GSS25-M5}). 

\medskip

With this background in hand, we turn to the corresponding analysis for the M-string.
  
\end{added}

%%%%%%%%%%%%%%%%%%%%%%%%%%%
\subsection{The Spin Geometry}
%%%%%%%%%%%%%%%%%%%%%%%%%%%%

Let $\Gamma_a$ denote the 11D Clifford generators satisfying (cf. \cite[\S 2.5]{MiemiecSchnakenburg2006}, we follow \cite[\S 2.2.1]{GSS24-SuGra})
\begin{equation}
  \label{TheCliffordRelations}
  \Gamma_a \Gamma_b
  +
  \Gamma_b \Gamma_a
  =
  +2 \, \eta_{ab}\, 
  \mathrm{id}_{\mathbf{32}}
  \mathrlap{\,,}
\end{equation}
where the tangential Minkowski metric is $\eta := \mathrm{diag}(-1,+1,\cdots, +1)$ and where $\mathbf{32}$ denotes the real 32-dimensional $\mathrm{Spin}(1,10)$-irrep with its $\mathrm{Spin}(1,10)$-invariant skew-symmetric bilinear form denoted
\begin{equation}
  \begin{tikzcd}[
    sep=20pt
  ]
  \bracket({
    \overline{(-)}
    (-)
  })
  :
  \mathbf{32}
    \otimes_{{}_{\mathbb{R}}} 
  \mathbf{32}
  \ar[r]
  &
  \mathbb{R}
  \mathrlap{\,,}
  \end{tikzcd}
\end{equation}
with respect to which $\overline{\Gamma_a} = - \Gamma_a$.  As usual, we denote the Clifford algebra basis elements as 
$$
  \Gamma_{a_1 \cdots a_p}
  :=
  \tfrac{1}{p!}
  \sum_{\sigma \in \mathrm{Sym}_p}
  \mathrm{sgn}(\sigma)
  \Gamma_{a_{\sigma(1)}}
  \Gamma_{a_{\sigma(2)}}
  \cdots
  \Gamma_{a_{\sigma(p)}}
$$
and we will use that (cf. \cite[Lem. 2.63]{GSS24-SuGra})
\begin{equation}
  \label{11DChiralityOperator}
  \Gamma_{0123455'6789}
  =
  \mathrm{id}_{\mathbf{32}}
  \mathrlap{\,,}
\end{equation}
reflecting the $\mathcal{N} = 1$ supersymmetry in 11D.

\begin{SCfigure}[1.15][htb]
\caption{\label{MStringSchematics}
 The \emph{M-string} (M1) is the transversal intersection of an M2-brane incident onto an M5-brane. In a Darboux frame adapted to the intersection, the tangent spaces of the M5 may be taken to extend along the 012345 axes, that of the M2 along 0125', hence that of the M1 along 01. 
 \begin{added}
 The configuration need not be rectangular as seen in the graphics, but in the superembedding Darboux coframe it does look like this in every tangent space.
 \end{added}
}
\centering
\adjustbox{
  rndfbox=4pt
}{
  \includegraphics[width=6.3cm]{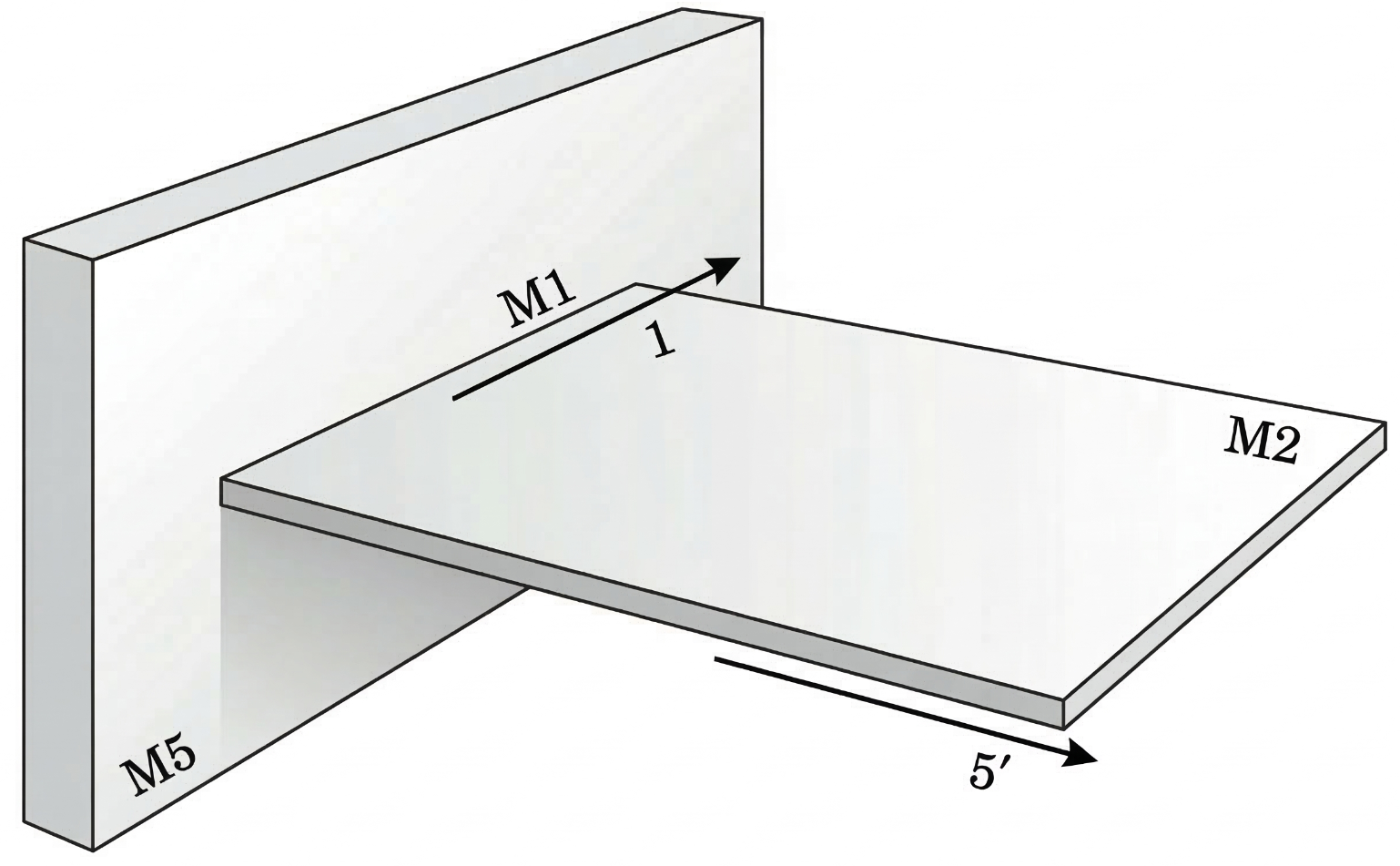}
}
\end{SCfigure}

Now, adapted to the tangent geometry of an M-string embedded into an M5-brane embedded into the 11D bulk (cf. \cref{MStringSchematics}), we decompose the tangential basis of Clifford generators like this:
\begin{equation}
  \begin{tikzcd}[
    column sep=-4pt,
    row sep=-2pt,
    /tikz/column 15/.append style={anchor=base west},
    execute at end picture={
      \draw[
        decorate, 
        decoration={
          brace, 
          amplitude=5pt, 
          raise=6pt
        }
      ]
        ([yshift=9pt]col0.north west) -- 
        ([yshift=9pt]col1.north east)
        node[midway, above=8pt] 
          {\scalebox{.7}{M-string}};
      \draw[
        decorate, 
        decoration={
          brace, 
          amplitude=5pt,
          aspect=.55
        }
      ]
        ([yshift=7pt]col0.north west) -- 
        ([yshift=7pt]col5.north east)
        node[pos=.55, above=4pt] 
          {\scalebox{.7}{M5-brane}};
      \draw[
        decorate, 
        decoration={
          brace, 
          amplitude=5pt,
          aspect=.5
        }
      ]
        ([yshift=6pt]col5prime.north west) -- 
        ([yshift=6pt]col5prime.north east)
        node[pos=.5, above=4pt] 
          {\scalebox{.7}{transverse}};
      \draw[
        decorate, 
        decoration={
          brace, 
          amplitude=5pt,
          aspect=.77
        }
      ]
        (col0.north west) -- 
        (col9.north east)
        node[pos=.77, above=4pt] 
          {\scalebox{.7}{bulk}};
    }
  ]
    |[alias=col0]|
    0 
    &
    |[alias=col1]|
    1
    &
    2
    &
    3
    &
    4
    &
    |[alias=col5]|
    5
    &
    |[alias=col5prime]|
    5'
    &
    6
    &
    7
    & 
    8
    & 
    |[alias=col9]|
    9
    \\
    \Gamma_0
    &
    \Gamma_1
    &
    \Gamma_2
    &
    \Gamma_3
    &
    \Gamma_4
    &
    \Gamma_5
    &
    \Gamma_{5'}
    &
    \Gamma_6
    &
    \Gamma_7
    &
    \Gamma_8
    &
    \Gamma_9
    &\in&
    \mathrm{Pin}^+\bracket({1,10})
    &\subset&
    \mathrm{End}_{{}_{\mathbb{R}}}
    \bracket({\mathbf{32}})
    \\
    \gamma_0 
    &
    \gamma_1
    & 
    \gamma_2
    & 
    \gamma_3
    & 
    \gamma_4
    &
    \gamma_5
    &&&&&
    &\in&
    \mathrm{Pin}^+\bracket({1,5})
    &\subset&
    \mathrm{End}_{{}_{\mathbb{R}}}
    \bracket({
      2\cdot \mathbf{8}_+ 
       \oplus 
      2 \cdot \mathbf{8}_-
    })
    \\
    \sigma_0
    &
    \sigma_1
    &&&&&&&&&
    &\in&
    \mathrm{Pin}^+\bracket({1,1})
    &\subset&
    \mathrm{End}_{{}_{\mathbb{R}}}
    \bracket({
       16 \cdot \mathbf{1}_+
       \oplus
       16 \cdot \mathbf{1}_-
    })
    \mathrlap{\,.}
  \end{tikzcd}
\end{equation}
Here the elements of the last two lines are defined by
\begin{subequations}
  \begin{align}
    P_{\!\!{}_\perp\!} \,\Gamma_a\, P
    +
    P \,\Gamma_a\, P_{\!\!{}_\perp\!}
    =
    \begin{cases}
      \gamma_a & \text{for $a$ tangential to M5}
      \\[-3pt]
      0 & \text{for $a$ transverse to M5}
    \end{cases}
    \\
    p \,\gamma_a\, p
    +
    p_{\!{}_\perp\!} \,\gamma_a\, p_{\!{}_\perp\!}
    =
    \begin{cases}
      \sigma_a & \text{for $a$ tangential to M1}
      \\[-3pt]
      0 & \text{for $a$ transverse to M1}
      \mathrlap{\,,}
    \end{cases}
  \end{align}
\end{subequations}
with respect to the following projection operators:
\begin{equation}
  \label{TheProjectionOperators}
  \begin{array}{clccl}
    P &:= 
    \tfrac{1}{2}\bracket({
      1 + \Gamma_{012345}
    }),
    &&
    P_{\!\!{}_\perp\!} &:= 
    \tfrac{1}{2}\bracket({
      1 - \Gamma_{012345}
    })
    \\
    p 
    &:=
    \tfrac{1}{2}\bracket({
      1 + \Gamma_{015'}
    })
    =
    \overline{p}
    ,
    &\hspace{1cm}&
    p_{\!{}_\perp\!} &:=
    \tfrac{1}{2}\bracket({
      1 - \Gamma_{015'}
    })
    \mathrlap{\,,}
  \end{array}
\end{equation}
satisfying
\begin{subequations}
\label{ProjectorBasicProperties}
\begin{align}
  \label{M5ProjectorBasicProperties}
  \begin{array}{ll}
    P P = P,
    &
    P_{\!\!{}_\perp\!} P = 0,
    \\
    P_{\!\!{}_\perp\!} \,P_{\!\!{}_\perp\!} = P_{\!\!{}_\perp\!},
    &
    P P_{\!\!{}_\perp\!} = 0
  \end{array}
  &\;\;\text{and}\;\;
  \left\{
  \begin{array}{ll}
    \Gamma_a P = P_{\!\!{}_\perp\!} \Gamma_a
    &
    \mbox{for $a$ tangential to M5},
    \\
    \Gamma_a P = P \Gamma_a
    &
    \mbox{for $a$ transverse to M5},
  \end{array}
  \right.
  \\
  \label{M1ProjectorBasicProperties}
  \begin{array}{ll}
    p \, p = p,
    &
    p_{\!{}_\perp\!} \, p = 0,
    \\
    p_{\!{}_\perp\!} \, p_{\!{}_\perp\!} = p_{\!{}_\perp\!},
    &
    p \, p_{\!{}_\perp\!} = 0
  \end{array}
  &\;\;\text{and}\;\;
  \left\{
  \begin{array}{ll}
    \gamma_a \, p  = p \, \gamma_a
    &
    \mbox{for $a$ tangential to M1},
    \\
    \Gamma_a \, p = p \, \Gamma_a
    &
    \mbox{for $a = 5'$},
    \\
    \Gamma_a \, p = p_{\!{}_\perp\!} \, \Gamma_a
    &
    \mbox{for $a \neq 5'$ transverse to M1},
  \end{array}
  \right.
\end{align}
\end{subequations}
\;\;
and
\;\;
\begin{equation}
  \begin{array}{ll}
    P \, p = p \, P
    \,,
    &   \hspace{1cm}
    P_{\!\!{}_\perp\!} \, p = 
    p \, P_{\!\!{}_\perp\!}
    \mathrlap{\,,}
    \\
    P \, p_{\!{}_\perp\!} = p_{\!{}_\perp\!} \, P
    \,,
    &   \hspace{1cm}
    P_{\!\!{}_\perp\!} \, p_{\!{}_\perp\!} = 
    p_{\!{}_\perp\!} \, P_{\!\!{}_\perp\!}
    \mathrlap{\,.}
  \end{array}
\end{equation}
The operator $P$ in \cref{TheProjectionOperators} projects onto that half of the spinors/susy charges which survives on the $\sfrac{1}{2}$-BPS M5-brane, while $p$ projects onto the further half of that which survives on an M2-brane intersecting the M5 locally in the 01-plane and extending transversally along the $5'$-axis. This is what characterizes the intersection as an \emph{M-string} (cf. \cite[(2.2)]{HaghighatEtAl2015}).

First, recall (following \cite[\S 3.2]{GSS25-M5}) the spinor representation on the M5-brane, projected out by the operator $P$. We observe that $P$ decomposes as
\begin{equation}
  \label{DecompositionOfP}
  \begin{aligned}
    P 
    & 
    \defneq
    \tfrac{1}{2}\bracket({
      1 + \Gamma_{012345}
    })
    =
    \tfrac{1}{2}\bracket({
      1 + \Gamma_{5'6789}
    })
    \\
    & 
    =
    \underbrace{
      \tfrac{1}{2}\bracket({
        1 + \Gamma_{5'}
      })
      \,
      \tfrac{1}{2}\bracket({
        1 + \Gamma_{6789}
      })
    }_{P_+}
    +
    \underbrace{
      \tfrac{1}{2}\bracket({
        1 - \Gamma_{5'}
      })
      \,
      \tfrac{1}{2}\bracket({
        1 - \Gamma_{6789}
      })
    }_{P_-}
    \mathrlap{\,,}
  \end{aligned}
\end{equation}
with the two summands exhibiting the irrep decomposition of the M5 spinor representation as \cite[(91)]{GSS25-M5}:
\begin{equation}
    P\bracket({\mathbf{32}})
    \underset{\mathrm{Spin}(1,5)}{\simeq}
    2\cdot \mathbf{8}_+
    \,,\;\;\;\;\;
    P_{\!\!{}_\perp\!}\bracket({\mathbf{32}})
    \underset{\mathrm{Spin}(1,5)}{\simeq}
    2\cdot \mathbf{8}_-
    \mathrlap{\,,}
\end{equation}
where $\mathbf{8}_{\pm}$ denote the two real 8-dimensional irreps of $\mathrm{Spin}(1,5)$. In addition, $P(\mathbf{32})$ and $P_{\!\!{}_\perp\!}(\mathbf{32})$ inherit a transversal $\mathrm{Spin}(5)$-action (acting in 5'6789, being the \emph{R-symmetry} group from the M5's worldvolume perspective). With respect to that, they are both isomorphic to the $4 \cdot \mathbf{4}$ of $\mathrm{Spin}(5)$ \cite[(98)]{GSS25-M5}:
$$
  2 \cdot \mathbf{8}_+
  \underset
    {\mathrm{Spin}(1,5)}
    {\simeq}
  P(\mathbf{32})
  \underset
    {\mathrm{Spin}(5)}
    {\simeq}  
  4 \cdot \mathbf{4} 
  \underset
    {\mathrm{Spin}(5)}
    {\simeq}  
  P_{\!\!{}_\perp\!}(\mathbf{32})
  \underset
    {\mathrm{Spin}(1,5)}
    {\simeq}  
  2 \cdot \mathbf{8}_-
  \mathrlap{\,.}
$$

Our task here is the analogous analysis of the induced spinor rep on the M-string. 
To that end, notice that, since the group $\mathrm{Spin}(1,1)$ is generated by exponentiating $\gamma_{01}$, it has two 1-dimensional spinorial irreps $\mathbf{1}_{\pm}$, characterized, as Clifford modules, by the two possible eigenvalues
\begin{equation}
  \gamma_{01}\vert_{\mathbf{1}_{\pm}}
  =
  \pm \mathrm{id}_{\mathbf{1}_{\pm}}
  \mathrlap{\,.}
\end{equation}
With this, we find that:
\begin{equation}
  \label{TheMStringSpinorRep}
  p P\bracket({\mathbf{32}})
  \underset{
    \mathrm{Spin}(1,1)
  }{\simeq}
  4 \cdot \mathbf{1}_+
  \oplus
  4 \cdot \mathbf{1}_-
  \underset{
    \mathrm{Spin}(1,1)
  }{\simeq}
  p_{\!{}_\perp\!} P(\mathbf{32})
  \mathrlap{\,,}
\end{equation}
For $p P(\mathbf{32})$ this may be seen as follows:
\begin{equation}
  \begin{aligned}
    \Gamma_{01}
    \,
    \vert_{pP(\mathbf{32})}
    & 
    \;
    \underset
      {\mathclap{\scalebox{.7}{
        \cref{TheProjectionOperators}
      }}}
      {=}
   \;
    \Gamma_{01}
    \Gamma_{015'}
    \Gamma_{012345}
    \,
    \vert_{pP(\mathbf{32})}
    \\
    & 
    \;
    \underset
      {\mathclap{\scalebox{.7}{
        \cref{TheCliffordRelations}
      }}}
      {=}
      \;
    \Gamma_{0123455'}
    \,
    \vert_{pP(\mathbf{32})}
    \\
    & 
    \;
    \underset
      {\mathclap{\scalebox{.7}{\cref{11DChiralityOperator}}}}
      {=}
      \;
    \Gamma_{6789}
    \,
    \vert_{pP(\mathbf{32})}
    \\
    & 
    \;
    \underset
      {\mathclap{\scalebox{.7}{\cref{DecompositionOfP}}}}
      {=}
      \;
    +
    \mathrm{id}
    \,
    \vert_{pP_+(\mathbf{32})}
    -
    \mathrm{id}
    \,
    \vert_{pP_-(\mathbf{32})}
    \mathrlap{\,,}
  \end{aligned}
\end{equation}
and analogously for $p_{\!{}_\perp\!} P(\mathbf{32})$:
\begin{equation}
  \begin{aligned}
    \Gamma_{01}
    \,
    \vert_{p_{\!{}_\perp\!} P(\mathbf{32})}
    &
    \;
    \underset
      {\mathclap{\scalebox{.7}{
        \cref{TheProjectionOperators}
      }}}
      {=}
      \;
    -
    \Gamma_{01}
    \Gamma_{015'}
    \Gamma_{012345}
    \,
    \vert_{p_{\!{}_\perp\!} P(\mathbf{32})}
    \\
    & = - \cdots
    \mathrlap{\,.}
  \end{aligned}
\end{equation}

\begin{SCfigure}[1][htb]
\caption{\label{TheGroupActions}
The presence of an M-string on an M5-brane breaks the local Lorentz spin symmetry from $\mathrm{Spin}(1,5)\times\mathrm{Spin}(5)$ to $\mathrm{Spin}(1,1) \times \mathrm{Spin}(4)_L \times \mathrm{Spin}(4)_R$.
}
\adjustbox{
  rndfbox=4pt
}{
$
\begin{tikzcd}[
  column sep=2pt,
    execute at end picture={
      \draw[
        decorate, 
        decoration={
          brace, 
          amplitude=5pt
        }
      ]
        (col0.north west) -- 
        (col1.north east)
        node[midway, above=2pt] 
          {$
            \overset
              {\mathrm{Spin}(1,1)}
              {
                \adjustbox{
                  rotate=190
                }{
                  $\circlearrowright$
                }
              }
          $};
      \draw[
        decorate, 
        decoration={
          brace, 
          amplitude=5pt
        }
      ]
        (col2.north west) -- 
        (col5.north east)
        node[pos=.5, above=2pt] 
          {$
            \overset
              {\mathrm{Spin}(4)_L}
              {
                \adjustbox{
                  rotate=190
                }{
                  $\circlearrowright$
                }
              }
          $};
      \draw[
        decorate, 
        decoration={
          brace, 
          amplitude=5pt
        }
      ]
        (col6.north west) -- 
        (col9.north east)
        node[pos=.5, above=2pt] 
          {$
            \overset
              {\mathrm{Spin}(4)_R}
              {
                \adjustbox{
                  rotate=190
                }{
                  $\circlearrowright$
                }
              }
          $};
    }
]
    |[alias=col0]|
    0 
    &
    |[alias=col1]|
    1
    &
    |[alias=col2]|
    2
    &
    3
    &
    4
    &
    |[alias=col5]|
    5
    &
    |[alias=col5prime]|
    5'
    &
    |[alias=col6]|
    6
    &
    7
    & 
    8
    & 
    |[alias=col9]|
    9
\end{tikzcd}
$
}
\end{SCfigure}
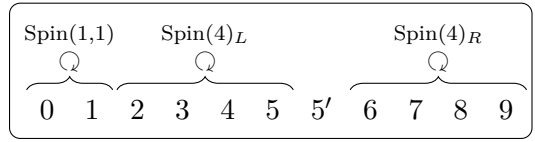

Moreover, the spinor subspaces $pP(\mathbf{32})$ and $p_{\!{}_\perp\!} P(\mathbf{32})$ inherit compatible representations of $\mathrm{Spin}(4)_L$ operating in 2345 and of $\mathrm{Spin}(4)_R$ operating in 6789 (cf. \cref{TheGroupActions}).

In order to analyze this transversal group action, observe that the projectors $p$ and $p_{\!{}_\perp\!}$ decompose as:
\begin{equation}
  \label{DecompositionOfp}
  \begin{aligned}
    p 
    & 
    =
    \tfrac{1}{2}\bracket({1 + \Gamma_{015'}})
    =
    \tfrac{1}{2}\bracket({ 1 + \Gamma_{23456789} })
    \\
    & 
    =
    \tfrac{1}{2}\bracket({1 + \Gamma_{2345}})
    \,
    \tfrac{1}{2}\bracket({1 + \Gamma_{6789}})
    -
    \tfrac{1}{2}\bracket({1 - \Gamma_{2345}})
    \,
    \tfrac{1}{2}\bracket({1 - \Gamma_{6789}})\,,
    \\[6pt]
    p_{\!{}_\perp\!} 
    & 
    =
    \tfrac{1}{2}\bracket({1 - \Gamma_{015'}})
    =
    \tfrac{1}{2}\bracket({ 1 - \Gamma_{23456789} })
    \\
    & 
    =
    \tfrac{1}{2}\bracket({1 - \Gamma_{2345}})
    \,
    \tfrac{1}{2}\bracket({1 + \Gamma_{6789}})
    -
    \tfrac{1}{2}\bracket({1 + \Gamma_{2345}})
    \,
    \tfrac{1}{2}\bracket({1 - \Gamma_{6789}})
    \mathrlap{\,,}
  \end{aligned}
\end{equation}
respectively.
Combined with \cref{DecompositionOfP} this yields:
\begin{equation}
  \label{DecompositionOfpP}
  \hspace{-3mm}
  \begin{aligned}
    p P
    & =
    \tfrac{1}{2}\bracket({1 + \Gamma_{2345}})
    \,
    \tfrac{1}{2}\bracket({1 + \Gamma_{5'}})
    \,
    \tfrac{1}{2}\bracket({1 + \Gamma_{6789}})
    +
    \tfrac{1}{2}\bracket({1 - \Gamma_{2345}})
    \,
    \tfrac{1}{2}\bracket({1 - \Gamma_{5'}})
    \,
    \tfrac{1}{2}\bracket({1 - \Gamma_{6789}})\,,
    \\[6pt]
    p_{\!{}_\perp\!} P
    & =
    \tfrac{1}{2}\bracket({1 - \Gamma_{2345}})
    \,
    \tfrac{1}{2}\bracket({1 + \Gamma_{5'}})
    \,
    \tfrac{1}{2}\bracket({1 + \Gamma_{6789}})
    +
    \tfrac{1}{2}\bracket({1 + \Gamma_{2345}})
    \,
    \tfrac{1}{2}\bracket({1 - \Gamma_{5'}})
    \,
    \tfrac{1}{2}\bracket({1 - \Gamma_{6789}})
    \mathrlap{\,.}
  \end{aligned}
\end{equation}

We see that $pP(\mathbf{32})$ is $\mathrm{Spin}(4)_L$-equivariantly isomorphic to $p_{\!{}_\perp\!}P(\mathbf{32})$, for instance via the action of $\Gamma_{5'6}$, using that $\Gamma_{5'6} \, p \, P = p_{\!{}_\perp\!} \, P \,\Gamma_{5'6}$ \cref{ProjectorBasicProperties}:
\begin{equation}
    \label{Spin4LIso}
    p P (\mathbf{32})
    \underset
      {\mathrm{Spin}(4)_L}
      {\simeq}
    p_{\!{}_\perp\!} P(\mathbf{32})
    \mathrlap{\,.}
\end{equation}
On the other hand, the two spinor spaces are \emph{not} $\mathrm{Spin}(4)_L \times \mathrm{Spin}(4)_R$-equivariantly isomorphic. This is because any such isomorphism has to commute with $\Gamma_{2345}\Gamma_{6789}$, while \cref{DecompositionOfpP} shows that this product operator acts differently on the two spaces:
\begin{equation}
  \label{pp32NotIsomorphicTopPerpP32}
    p P (\mathbf{32})
    \underset
      {
        \mathrm{Spin}(4)_L 
          \times 
        \mathrm{Spin}(4)_R
      }
      {\not\simeq}
    p_{\!{}_\perp\!} P(\mathbf{32})
    \mathrlap{\,.}
\end{equation}
More concretely, with 
$$
  \pi
  :=
  \tfrac{1}{2}\bracket({
    1 + \Gamma_{2345}
  })
  \,,
  \;\;\;
  \pi'
  :=
  \tfrac{1}{2}\bracket({
    1 - \Gamma_{2345}
  })\,,
$$
we have
\begin{equation}
  \begin{aligned}
  \mathrm{Spin}(4)_L
  & \simeq
  \mathrm{Spin}(3) 
    \times 
  \mathrm{Spin}(3)
  \\
  & \simeq
  \exp\bracket({
    \bracket\langle{
      \pi \Gamma_{23}
      ,
      \pi \Gamma_{24}
      ,
      \pi \Gamma_{34}
     }\rangle
   })
   \exp\bracket({
    \bracket\langle{
      \pi' \Gamma_{23}
      ,
      \pi' \Gamma_{24}
      ,
      \pi' \Gamma_{34}
    }\rangle
  })
  \mathrlap{\,,}
  \end{aligned}
\end{equation}
whence $\mathrm{Spin}(4)_L$-representations on which $\Gamma_{2345}$ has positive or negative eigenvalue are left-chiral, $(\mathbf{n},\mathbf{1})$, or right-chiral, $(\mathbf{1},\mathbf{n})$, respectively, as representations of $\mathrm{Spin}(3) \times \mathrm{Spin}(3)$.
But since the analogous statement holds also for $\mathrm{Spin}(4)_R$, it follows with \cref{DecompositionOfpP} (using also $\Gamma_{5'}\vert_{pP(\mathbf{32})} = \Gamma_{01}\vert_{p P (\mathbf{32})}$ and $\Gamma_{5'}\vert_{p_{\!{}_\perp\!}P(\mathbf{32})} = -\Gamma_{01}\vert_{p_{\!{}_\perp\!} P (\mathbf{32})}$)
that (for the first line, cf. \cite[(2.4)]{HaghighatEtAl2015}):
\begin{equation}
\hspace{-4mm} 
  \begin{aligned}
  p P(\mathbf{32})
  &
  \underset
    {
      \mathrm{Spin}(1,1)
        \times
      \mathrm{Spin}(4)_L 
        \times 
      \mathrm{Spin}(4)_R
    }
    {\simeq}
  \mathbf{1}_+ 
  \boxtimes
  (\mathbf{2},\mathbf{1})
  \boxtimes 
  (\mathbf{2},\mathbf{1})
  \;\oplus\;
  \mathbf{1}_- 
  \boxtimes
  (\mathbf{1},\mathbf{2})
  \boxtimes 
  (\mathbf{1},\mathbf{2})\,,
  \\
  p_{\!{}_\perp\!} P(\mathbf{32})
  &
  \underset
    {
      \mathrm{Spin}(1,1)
        \times
      \mathrm{Spin}(4)_L 
        \times 
      \mathrm{Spin}(4)_R
    }
    {\simeq}
  \mathbf{1}_+ 
  \boxtimes
  (\mathbf{2},\mathbf{1})
  \boxtimes 
  (\mathbf{1},\mathbf{2})
  \;\oplus\;
  \mathbf{1}_- 
  \boxtimes
  (\mathbf{1},\mathbf{2})
  \boxtimes 
  (\mathbf{2},\mathbf{1})
  \mathrlap{\,.}
  \end{aligned}
\end{equation}
In particular, this confirms \cref{pp32NotIsomorphicTopPerpP32}, which is the relevant statement for the following superembedding construction.

%%%%%%%%%%%%%%%%%%%%%%%%%%%%%%%%%%%%%%%
\subsection{The Iterated Embedding}
\label{OnTheIteratedEmbedding}
%%%%%%%%%%%%%%%%%%%%%%%%%%%%%%%%%%%%%%%

Given an 11D SuGra spacetime $X^{1,10}$,
we consider a sequence of $\sfrac{1}{2}$-BPS superembeddings \cite[\S 2.2]{GSS25-M5}: first of a super M5 worldvolume into the 11D bulk, and then of a string worldsheet into the M5. We use the following notation to describe this situation (cf. again \cref{MStringSchematics}):
\begin{equation}
  \label{TheSuperembeddings}
  \hspace{-1cm}
  \begin{tikzcd}[
    row sep=-2pt, 
    column sep=20pt
  ]
    &[-25pt] 
    \substack{
      \text{\color{darkblue} \bf M-string}
    }
    &
    \substack{
      \text{\color{darkblue} \bf M5-brane}
    }
    &
    \substack{
      \text{\color{darkblue} \bf 11D bulk}
    }
    \\
    \substack{
      \text{\color{darkblue}Worldvolume}
    }
    &
    N^{1,1}
    \;
    \ar[
      r, 
      hook,
      "{ \phi }"
    ]
    &
    \quad 
    \Sigma^{1,5}
    \quad 
    \ar[
      r, 
      hook,
      "{ \Phi }"
    ]
    &
    \;
    X^{1,10}
    \\
    \substack{
      \text{\color{darkblue}Super-coframe}
    }
    &
    (\StringVielbein,\StringGravitino)
    &
    (e,\psi)
    &
    (E,\Psi)
    \\
    \substack{
      \text{\color{darkblue}Spin connection}
    }
    &
    \varpi
    &
    \omega
    & 
    \Omega
    \\
    \substack{
      \text{\color{darkblue}Clifford algebra}
    }
    &
    \sigma & \gamma & \Gamma
    \mathrlap{\,.}
  \end{tikzcd}
\end{equation}
Since our goal is the analysis of Gauss laws/Bianchi identities for flux densities on these objects, it will be sufficient to determine the torsion constraints, which we obtain as \cref{TorsionConstraintOnM5,TorsionConstraintOnString} below. However, to determine this, we first need to analyze the iterated superembedding in more detail.

In \cref{TheSuperembeddings}, the coframes are understood to be adapted to the superembeddings (\emph{Darboux coframes}, cf. \cite[\S 2.1]{GSS25-M5}) in that 
\begin{equation}
  \label{DarbouxCoframeCondition}
  \phi^\ast e^a 
    \!=\!
  \begin{cases}
    \StringVielbein^a & \text{for $a$ tangential to M1}
    \\[-2pt]
    0 & \text{for $a$ transversal to M1}
  \end{cases}
  \,,\quad 
  \Phi^\ast E^a 
    \!=\!
  \begin{cases}
    e^a & \text{for $a$ tangential to M5}
    \\[-2pt]
    0 & \text{for $a$ transversal to M5.}
  \end{cases}
\end{equation}
But the supergeometric generalization of this Darboux condition allows, \emph{a priori}, for an odd component of the superembedding to involve \emph{fermionic shear} operators (\cite[(15)]{HoweSezgin1997}, cf. \cite[(106)]{GSS25-M5}),
\begin{equation}
  \label{TheShearOperators}
  \slashed{\tilde H}_1
    :=
  (\tilde H_1)_a \sigma^a
  \Gamma_{5'6}
  \,,
  \;\;\;
  \slashed{\tilde H}_3 
    := 
  \tfrac{1}{3}(\tilde H_3)_{a_1 a_2 a_3} \gamma^{a_1 a_2 a_3}
  \mathrlap{\,,}
\end{equation}
appearing in the pullback of the ambient fermion coordinates like this:
\begin{subequations}
  \label{OddComponentsOfSuperembeddings}
  \begin{align}
  &
  \begin{tikzcd}[
    ampersand replacement=\&,
    row sep=-1pt,
    column sep=30
  ]
    \&\&
    \hspace{38pt}
    \overbrace
      {\phantom{-------}}
      ^{ \mathbf{32} }
    \&
    \hspace{.5cm}
    \\[-15pt]
    \&\&
      2 \cdot \mathbf{8}_+
      \mathrlap{
        \,\oplus\,\,
        2 \cdot \mathbf{8}_-
      }
    \ar[d, ->>]
    \\[20pt]
    T \Sigma
    \ar[
      rr,
      "{
        \psi
      }",
      "{
        \scalebox{.7}{odd coframe on M5}
      }"{swap}
    ]
    \ar[
      urr,
      bend left=4pt,
      "{
        \Phi^\ast \Psi
        \,=\,
        \bracket({
          \psi
          ,
          \slashH_3 
          \psi
        })
      }"{sloped}
    ]
    \&\&
    2 \cdot \mathbf{8}_+
    \mathrlap{\,,}
  \end{tikzcd}
  \\
  &
  \begin{tikzcd}[
    ampersand replacement=\&,
    row sep=-1pt,
    column sep=38
  ]
    \&\&
    \mathrlap{
    \hspace{44pt}
    \mathclap{
      \overbrace
        {\phantom{----------------}}
        ^{ 2 \cdot \mathbf{8}_+ }
      }
    }
    \& \hspace{.5cm}
    \\[-15pt]
    \&\&
      (4 \cdot \mathbf{1}_+
      \oplus 
      4 \cdot \mathbf{1}_-)
      \mathrlap{
        \,\oplus\,\,
      (4 \cdot \mathbf{1}_-
      \oplus 
      4 \cdot \mathbf{1}_+)
      }
    \ar[d, ->>]
    \\[20pt]
    T N
    \ar[
      rr,
      "{
        \StringGravitino
      }",
      "{
        \scalebox{.7}{odd coframe on M1}
      }"{swap}
    ]
    \ar[
      urr,
      bend left=4pt,
      "{
        \phi^\ast \psi
        \,=\,
        \bracket({
          \StringGravitino
          ,
          \slashH_1 
          \StringGravitino
        })
      \;\;}"{sloped}
    ]
    \&\&
    4 \cdot \mathbf{1}_+
    \oplus 
    4 \cdot \mathbf{1}_-
    \mathrlap{\,.}
  \end{tikzcd}
  \end{align}
\end{subequations}
The shear operator $\tilde H_3$ in \cref{TheShearOperators}, which is necessarily self-dual (cf. \cite[Lem. 3.8]{GSS25-M5}), is the origin of the non-linearly self-dual 3-form flux $H_3$ on the M5 (cf. \cite[Prop. 3.18, Rm. 3.19]{GSS25-M5} and \cref{FluxQuantizingOnM5TheM3Flux} below). Its square  (cf. \cite[(70)]{GSS25-M5})
\begin{equation}
  \bracket({{\tilde H_3}{}^2})^b_a := \bracket({\tilde H_3})_{a c_1 c_2} \bracket({\tilde H_3})^{b c_1 c_2}
\end{equation}
will play a key role shortly \eqref{TorsionConstraintOnM5}. 

The shear operator $\slashed{\tilde H}_1$ in \cref{TheShearOperators} is superficially an analogous source of a 1-form flux on the M-string (the factor $\Gamma_{5'6}$ is needed to make it actually shear into the orthogonal spin representation, by above \cref{Spin4LIso}). We will see shortly \cref{TildeH1Vanishes} that $\slashed{\tilde{H}}_1$ is dynamically constrained to vanish --- and the remainder of the article, in \cref{OnProperFluxQuantization} and \cref{OnMStringFluxQuantization}, is concerned with explaining how vanishing M-brane flux may still have topologically nontrivial and noteworthy gauge potentials.
%%%%%%%%%%%%%%%%%%%%%%%%%%%%%%%%%%%%%%%%%%%%%%%%
\label{OnTheEquivarianceOfTheM1ShearOperator}
%%%%%%%%%%%%%%%%%%%%%%%%%%%%%%%%%%%%%%%%%%%%%%%%

However, we want to highlight that $\slashed{\tilde H}_1$ should be taken to vanish already on more fundamental grounds:
Namely, the conceptual origin of the fermionic shears \cref{OddComponentsOfSuperembeddings} among super-embeddings is (cf. \cite[Rem. 2.22]{GSS25-M5}) the possibility that half of the spinors/susy charges which survive on the embedded brane is \emph{equivariantly} isomorphic to the broken orthogonal half, with respect to the transversal Lorentz spin symmetry.

This traditional rule of superembedding needs refinement in our case of iterated superembeddings: While $\slashed{\tilde H}_1$ in \cref{OddComponentsOfSuperembeddings} is $\mathrm{Spin}(4)_L$-equivariant \cref{Spin4LIso}, as befits a string superembedding into an 5-brane \emph{seen in isolation}, it is \emph{not} \cref{pp32NotIsomorphicTopPerpP32} equivariant with respect to the actual larger transversal symmetry group $\mathrm{Spin}(4)_L \times \mathrm{Spin}(4)_R$, which takes into account that the M5-brane itself is embedded into a larger bulk (cf. \cref{TheGroupActions}).

So if we demand, as appears reasonable, that the rule for \emph{iterated superembeddings} must be their equivariance under the full iterated transversal symmetry group, then $\slashed{\tilde H}_1$ in \cref{TheShearOperators} must vanish already on these grounds, and with it also any other shear component $\slashed{\tilde H}_0$ and $\slashed{\tilde H}_2$, which could otherwise be present for an M-string embedding.

%%%%%%%%%%%%%%%%%%%%%%%%%%%%%%%%%%%%%
\subsection{The Torsion Constraint}
\label{OnTheTorsionConstraint}
%%%%%%%%%%%%%%%%%%%%%%%%%%%%%%%%%%%%%

The dynamics of supergravity is largely controlled by \emph{torsion constraints} (cf. \cite{Lott1990}). 
In 11D, this constraint just says that the bosonic components of the \emph{super-}torsion tensor of the super-spin geometry $(E,\Psi,\Omega)$ in \cref{TheSuperembeddings} vanish (cf. \cite{Howe1997}):
\begin{equation}
  \label{11DSuGraTorsionConstraint}
  \mathrm{d}E^a 
    + 
  \Omega^a{}_b E^b
  -
  \bracket({\,
    \overline{\Psi}
    \Gamma^a
    \Psi
  })
  = 
  0
  \mathrlap{\,.}
\end{equation}
Indeed, the equations of motion of 11D SuGra are already equivalent (\cite[Thm. 3.1]{GSS24-SuGra}) to this super-torsion constraint combined with the Gauss/Bianchi identity on the super-flux densities \cref{OnShell11DSuGra}, we come back to this remarkable phenomenon in \S\ref{OnSuperFluxQuantization} below.

Pulling back the 11D torsion constraint
\eqref{11DSuGraTorsionConstraint} along the super-embedding $\Phi$ \cref{TheSuperembeddings}, of an M5-brane probe with a fermionic shear $\slashed{\tilde H}_3$ \cref{OddComponentsOfSuperembeddings}, yields (among other equations of motion, cf. \cite[\S 5.2]{Sorokin2000}) the \emph{torsion constraint on the M5-brane} \cite[(116)]{GSS25-M5}:
\begin{equation}
  \label{TorsionConstraintOnM5}
  \mathrm{d}\,
  e^a
  +
  \omega^a{}_b\, e^b
  =
  \CoefficientMatrix^a_{a'}
  \bracket({\,
    \overline{\psi}
    \,\gamma^{a'}\,
    \psi
  })
  \,,
  \;\;
  \text{where } \;
  \CoefficientMatrix^a_{a'}
   :=
  \bracket({
    \delta^{a}_{a'}
    -
    2({
      \tilde H_3{}^2
    })^a_{a'}
  })
  \mathrlap{\,.}
\end{equation}

In this vein, the further pullback of this M5-torsion constraint \eqref{TorsionConstraintOnM5} along an M-string super-embedding $\phi$ \cref{TheSuperembeddings} yields the equations of motion of the M-string. Here we focus on the following two components:

\begin{enumerate}
\item
Since the pullback of the M1-transversal component of \cref{TorsionConstraintOnM5} along $\phi$ has no $(\psi^2)$-terms on the left, by \cref{DarbouxCoframeCondition}, specifically the pullback of the $5'$ component of \cref{TorsionConstraintOnM5} gives the following condition, by \cref{M5ProjectorBasicProperties} and as in \cite[(118)]{GSS25-M5}:
\footnote{
  The other transverse components with $a' \in \{2,3,4,5,6,7,8,9\}$ give no further constraints, since for them the term analogous to \eqref{DerivingVanishingOfTildeH1} vanishes identically, now using the properties of $p$ \eqref{M1ProjectorBasicProperties}:
  $$
   \begin{aligned}
     \phi^\ast
     \bracket({\,
       \overline{\psi}
       \gamma^{a'}
       \psi
     })
     &=
     \bracket({
       \overline{\StringGravitino}
       \,
       p
       \,
       \bracket({%
         1
         -
         \slashed{\tilde H}_1
       })
       \,
       \gamma^{a'}
       \,
       \bracket({
         1
         +
         \slashed{\tilde H}_1
       })
       p
       \,
       \StringGravitino
     })
     \\
     & =
     \bracket({
       \overline{\StringGravitino}
       \,
       p p_{\!{}_\perp\!}
       \,
       \bracket({%
         1
         -
         \slashed{\tilde H}_1
       })
       \,
       \gamma^{a'}
       \,
       \bracket({
         1
         +
         \slashed{\tilde H}_1
       })
       \,
       \StringGravitino
     })
     = 0
     \mathrlap{\,.}
   \end{aligned}
  $$
}
 \begin{equation}
   \label{DerivingVanishingOfTildeH1}
   \begin{aligned}
     0 
     & =
     \phi^\ast
     \bracket({
       \overline{\psi}
       \gamma^{5'}
       \psi
     })
     \\
     &=
     \bracket({
       \overline{\StringGravitino}
       \,
       P_{\!\!{}_\perp\!}
       \,
       \bracket({%
         1
         -
         \slashed{\tilde H}_1
       })
       \,
       \gamma^{5'}
       \,
       \bracket({
         1
         +
         \slashed{\tilde H}_1
       })
       P
       \,
       \StringGravitino
     })
     \\
     & =
     2
     \bracket({
       \overline{\StringGravitino}
       \,
         \slashed{\tilde H}_1
         \gamma^{5'}
       \,
       \StringGravitino
     })
     \mathrlap{\,.}
   \end{aligned}
\end{equation}
This means that (cf. the discussion on p. \pageref{OnTheEquivarianceOfTheM1ShearOperator}):
\begin{equation}
  \label{TildeH1Vanishes}
  \slashed{\tilde H}_1 \defneq 0
  \mathrlap{\,.}
\end{equation}

\item 
With that, the pullback of the M1-tangential component of \cref{TorsionConstraintOnM5} immediately gives the \emph{torsion constraint on the M-string}:
\begin{equation}
  \label{TorsionConstraintOnString}
  \mathrm{d}\,
  \StringVielbein^a
  +
  \StringSpinConnection^a{}_b\, 
  \StringVielbein^b
  =
  \bracket({
    \phi^\ast
    \CoefficientMatrix
  })^a_{a'}
  \bracket({
    \overline{\StringGravitino}
    \,\sigma^{a'}\,
    \StringGravitino
  })
  \mathrlap{\,.}
\end{equation}
(Here all indices are in $\{0,1\}$, including those being summed over.) 
\end{enumerate}

We shall assume in the following that the 3-flux density on the M5-brane is non-critical on the M-string, meaning that the coefficient matrix on the right of \cref{TorsionConstraintOnString} is invertible:
\begin{equation}
  \label{NonCriticalityCondition}
  \mathrm{det}
  \bracket({
    \phi^\ast
    \CoefficientMatrix
  })
  \overset{!}{\neq}
  0
    \mathrlap{\,.}
\end{equation}
This is the case on an open neighborhood around $\tilde H_3 = 0$, hence in particular in the commonly considered \emph{small field limit} in which the corresponding 3-flux $H_3$ density (which is a quadratic function of $\tilde H_3$) is approximately Hodge self-dual, constituting the famous self-dual tensor field in the $D=6$, $\mathcal{N}= (2,0)$ worldvolume SCFT (cf. \cite[Rem. 3.19]{GSS25-M5}).

Beyond \cref{TildeH1Vanishes,TorsionConstraintOnString},
there are further constraints implied by the pullback of the torsion constraint \cref{TorsionConstraintOnM5} along the M-string superembedding, and these translate to the equations of motion of the M-string inside the M5-brane. Here, we do not further dwell on this local dynamics but turn attention now to global topological aspects that have not previously found the attention they deserve.

Crucially, we want to highlight that if with \cref{TildeH1Vanishes} we identify a 1-form flux density on the M-string worldvolume which is \emph{on-shell vanishing},
\begin{equation}
  \label{H1}
  H_1 \in \Omega^1_{\mathrm{dR}}\bracket({
    N^{1,1}
  })
  \,,
  \;\;\;\;
  H_1 = 0
  \mathrlap{\,,}
\end{equation}
then this is \emph{only locally} redundant as field content. Globally, there may be nontrivial topological quantum observables associated with the global completion of the $H_1$ field.
This phenomenon is familiar from (abelian) Chern-Simons theory (whose flux density $F_2$ also vanishes on-shell, $F_2 = 0$), but it may remain underappreciated in the context of M-branes. 

Our next goal is to work out a topological global completion of the field content of 11D SuGra probed by M5-branes probed by M-strings, including the 1-form flux \cref{H1}.

%%%%%%%%%%%%%%%%%%%%%%%%%
\section{Proper Flux Quantization}
\label
{OnProperFluxQuantization}
%%%%%%%%%%%%%%%%%%%%%%%%%

Beyond local equations of motion on flux densities,
the global definition of (higher) gauge fields requires a choice of \emph{flux/charge quantization} law \cite{SS25-Flux}. This is in principle well-known, but in practice most discussions still focus on just the field content visible on a single chart of spacetime. We highlight the rich space of choices involved in defining a theory's field content globally.

In the generality that we need here, which includes electric fluxes satisfying non-linear Gauss laws, flux quantization takes place not just in Whitehead-generalized cohomology (such as ordinary cohomology or K-theory, cf. \cite{Freed2000}) but in \emph{nonabelian cohomology} \cite{SS25-Flux,SS24-Phase,FSS23-Char}.

We now give a brief review, leading from Dirac's original insight into charge quantization in Maxwell theory to the proper flux quantization of magnetized M5-brane probes, which is the starting point for flux quantization on the M-string in \cref{OnMStringFluxQuantization} below.

Aimed at a theoretical physics audience, the following assumes some familiarity with basic concepts of algebraic topology, but will sacrifice mathematical fine print for ease of readability.

%%%%%%%%%%%%%%%%%%%%%%%%%%%%%%%
\subsection{Flux Quantization}
%%%%%%%%%%%%%%%%%%%%%%%%%%%%%%%

The notion of flux/charge quantization goes back to Dirac's foundational observation that, in modern paraphrase, the ordinary magnetic Gauss law
\begin{equation}
  \mathrm{d} F_2
  = 0
  \;\;\;\;
  {\color{gray}\substack{
    \text{magnetic}
    \\
    \text{Gauss law}
  }}
\end{equation}
entails that globally the electromagnetic field consists, beyond its \emph{flux density} $F_2$, also of a map from spacetime $X^{1,3}$ to a \emph{classifying space}, like 
\begin{equation}
  \label{ClassifyingOrdinaryCharge}
  \begin{tikzcd}
    X^{1,3} 
    \ar[r, dashed]
    &
    B \mathrm{U}(1) 
    \simeq
    K\bracket({\mathbb{Z},2})
  \;\;\;\;
  {\color{gray}\substack{
    \text{global}
    \\
    \text{charge}
  }}
    \mathrlap{\,,}
  \end{tikzcd}
\end{equation}
representing a class in the ordinary cohomology $H^2\bracket({X^{1,3};\mathbb{Z}})$: the \emph{magnetic charge} which is sourcing the magnetic flux.

More recently it is understood that also the vacuum electric Gauss law
$
  \mathrm{d} F'_2 = 0
$
(for $F'_2 := \star_4 F_2$) is potentially quantized, an admissible classifying space being the product space $K\bracket({\mathbb{Z},2}) \times K\bracket({\mathbb{Z},2})$. 

\textbf{The general rule for flux quantization} is essentially the following: 
\footnote{
  The math behind \cref{GenericGaussLaw} is that of \emph{minimal Sullivan models} in rational homotopy theory (cf. \cite[\S 5]{FSS23-Char}, as reviewed in \cite[\S 3.1]{SS25-Flux}), closely related to the ``FDAs'' of the SuGra literature (cf. \cite{FSS19-RationalM}).
}
Given  a set of (electric \& magnetic) flux densities $F^{(i)}$ (differential forms of some degree $\mathrm{deg}(i)$) satisfying Gauss/Bianchi identities
\begin{equation}
  \label{GenericGaussLaw}
  \mathrm{d}
  F^{(i)}
  =
  P^{(i)}\bracket({
    F^{(1)}, F^{(2)}, \cdots
  })
\end{equation}
for graded-symmetric polynomials $P^{(i)}$, then the admissible classifying spaces $\mathcal{A}$ are those whose real cohomology is computed by the cohomology of these differential relations when the $F^{(i)}$ are regarded as abstract algebra generators $f^{(i)}$:
\begin{equation}
  \label{GeneralRuleForFluxQuantization}
  \hspace{-5mm} 
  {\color{gray}\substack{
    \text{Admissible}
    \\
    \text{global charge}
  }}
  \;\;
  \begin{tikzcd}[sep=small]
    X
    \ar[r, dashed]
    &
    \mathcal{A}
  \end{tikzcd}
  \;\;
  \text{if}
  \;\;
  H^\bullet\bracket({
    \mathcal{A}
  })
  \overset{!}{\simeq}
  \frac
    {\mathrm{ker}(\mathrm{d})}
    {\mathrm{im}(\mathrm{d})}
  \,,\;\;
  \text{where}
  \;
  \mathrm{d}f^{(i)}
  =
  P^{(i)}\bracket({f^{(1)}, f^{(2)}, \cdots})
  \mathrlap{\,.}
\end{equation}

For example, the real cohomology of the higher classifying space $B^n \mathrm{U}(1)$ (the \emph{Eilenberg-MacLane space} $K(\mathbb{Z},n+1)$) is the graded polynomial algebra on single closed generator $f_{n+1}$, $\mathrm{d} f_{n+1} = 0$; whence this is the standard candidate classifying space for plain $n$-form gauge fields:
\begin{equation}
  \label{QuantizationOfHigherMagneticFlux}
  {\color{gray}\substack{
    \text{magnetic}
    \\
    \text{Gauss law}
  }}
  \;\;\;\;
  \mathrm{d} F_{n+1} 
  = 0
  \,,
  \;\;\;\;\;\;
  \begin{tikzcd}
    X^{1,d}
    \ar[r, dashed]
    &
    K\bracket({\mathbb{Z}, n+1})
  \end{tikzcd}
  \;\;\;\;
  {\color{gray}\substack{
    \text{global}
    \\
    \text{charge}
  }}
  \mathrlap{\,.}
\end{equation}
This gets doubled when also the electric fluxes $F'_{d-n} := \star F_{n+1}$ are quantized:
\begin{equation}
  \label{FluxQuantizingHigherEMFields}
  {\color{gray}\substack{
    \text{electro/magnetic}
    \\
    \text{Gauss laws}
  }}
  \;\;\;\;
  \begin{aligned}
    \mathrm{d} F_{n+1} 
    & = 0
    \\
    \mathrm{d} F'_{d-n} 
    & = 0    
    \,,
  \end{aligned}
  \;\;\;\;\;\;
  \begin{tikzcd}
    X^{1,d}
    \ar[r, dashed]
    &
    \begin{aligned}
     K\bracket({\mathbb{Z}, n+1})
     \\
     \times 
     K\bracket({\mathbb{Z},d-n})
    \end{aligned}
  \end{tikzcd}
  \;\;\;\;
  {\color{gray}\substack{
    \text{global}
    \\
    \text{charge}
  }}
  \mathrlap{\,.}
\end{equation}
For another example, the real cohomology of the classifying space $\mathrm{KU}_0 := \cup_{n} \mathrm{B} U(n) \times \mathbb{Z}$ for \emph{K-theory} is the graded polynomial algebra on closed generators $f_{2\bullet}$, $\mathrm{d} f_{2\bullet} = 0$ in all even degrees (the \emph{Chern classes}); whence this is a famously conjectured candidate classifying space for the RR-field in type IIA supergravity (cf. \cite[\S 4.1]{SS25-Flux}):
\begin{equation}
  \label{ChargeQuantizationInKTheory}
  {\color{gray}\substack{
    \text{electro/magnetic}
    \\
    \text{Gauss laws}
  }}
  \;\;\;\;
  \mathrm{d} F_{2\bullet} = 0
  \,,
  \;\;\;\;\;\;
  \begin{tikzcd}
    X^{1,d}
    \ar[r, dashed]
    &
    \mathrm{KU}
  \end{tikzcd}
  \;\;\;\;
  {\color{gray}\substack{
    \text{global}
    \\
    \text{charge}
  }}
  \mathrlap{\,.}
\end{equation}
Notice how in this case the electric flux densities $F_{2\bullet > 5}$ are quantized by the same classifying space as the magnetic flux densities $F_{2\bullet < 5}$, not splitting off via a product space as in \cref{FluxQuantizingHigherEMFields}.

So far, these examples all involve vanishing differentials, hence \emph{linear} Gauss laws whose solutions form a vector space. This is the case when the classifying space is an (infinite) \emph{loop space} like a stage in a \emph{spectrum of spaces} representing a Whitehead-generalized abelian cohomology theory like K-theory, elliptic cohomology or stable cobordism. Flux quantization in this linear/abelian case has been discussed in \cite{Freed2000} (cf. also \cite{SS23-Mf}).

However, this is not the general case. 
Notably the Gauss law for the electric flux $G_7 := \star G_4$ in 11D supergravity is famously quadratic,
$
  \mathrm{d}\,
    \mathrm{G}_7
    =
    \tfrac{1}{2}
    G_4 \wedge G_4
$.
This means that the common idea that the magnetic 4-flux $G_4$ can be quantized in $K\bracket({\mathbb{Z},4})$ (and be it a ``shifted'' version of that as proposed in \cite{HopkinsSinger2005}, cf. \cite[Prop. 3.1.1]{FSS15-ModuliStack}), as in \cref{QuantizationOfHigherMagneticFlux}, necessarily fails to account for any electric flux quantization as in \cref{FluxQuantizingHigherEMFields}. 

But from \cref{GeneralRuleForFluxQuantization} one sees how to fix this: We need a variant $\mathcal{A}$ of $K\bracket({\mathbb{Z},4})$ which still carries a cohomology class $f_4$, but the square $f_4^2$ of which vanishes in real cohomology. This is well-known (cf. \cite[\S 1.2]{Menichi2015}\cite[Ex. 5.3]{FSS23-Char}\cite[(17,27)]{SS25-Flux}) to be the case on the 4-sphere $S^4 \subset K\bracket({\mathbb{Z},4})$ which hence is a candidate classifying space for C-field charge (cf. \cite[\S 2]{FSS17-Sphere}\cite[\S 2.5]{Sati2018}\cite[\S 7]{FSS19-RationalM}):
\begin{equation}
  \label{FluxQuantizing11DSuGraEMField}
  {\color{gray}\substack{
    \text{electro/magnetic}
    \\
    \text{Gauss laws}
  }}
  \;\;\;\;
  \begin{aligned}
    \mathrm{d}\, G_7 & = 
    \tfrac{1}{2} G_4 \wedge G_4
    \\
    \mathrm{d}\, G_4 & = 0
    \,,
  \end{aligned}
  \;\;\;\;\;\;
  \begin{tikzcd}
    X^{1,d}
    \ar[r, dashed]
    &
    S^4
  \end{tikzcd}
  \;\;\;\;
  {\color{gray}\substack{
    \text{global}
    \\
    \text{charge}
  }}
  \mathrlap{\,.}
\end{equation}

As with all flux quantizations, there are infinitely many other choices one could make for $\mathcal{A}$. For instance, with any finite group $K$, also $\mathcal{A} := S^4 \times B K$ satisfies the condition \cref{GeneralRuleForFluxQuantization}. But $S^4$ is in a precise sense the minimal choice of classifying space (having a single ``cell'') for 11D SuGra. 

Since the nonabelian cohomology theory classified by spheres is called \emph{co-Homotopy} (\cite[\S VII]{STHu59}\cite[Ex. 2.7]{FSS23-Char}, dual to the \emph{homotopy} groups of maps out of spheres), the hypothesis that this is the ``correct'' choice for the global completion of 11D SuGra has been called \emph{Hypothesis H} \cite{FSS20-H,FSS21-Hopf,SS23-Mf}. This hypothesis is supported by how it provably implies subtle topological effects that are expected in the completion of 11D SuGra to ``M-theory''. In particular, the tangentially twisted version of 4-Cohomotopy \eqref{FluxQuantizing11DSuGraEMField} (which we disregard here just for brevity) does imply \cite[Prop. 3.13]{FSS20-H} the subtle half-integral shifting (by $\tfrac{1}{4}p_1$) of the de Rham class of $G_4$ that had famously been argued for in \cite[(1.2)]{Witten1997flux}.

%%%%%%%%%%%%%%%%%%%%%%%%%%%%%%%%%%%%%%%%%%%
\subsection{Relative Flux Quantization}
%%%%%%%%%%%%%%%%%%%%%%%%%%%%%%%%%%%%%%%%%%%

More generally, Gauss laws hold \emph{relatively}. 
For instance, in general the ordinary electric Gauss law for $F'_2 := \star F_2$ is of course
\begin{equation}
  \label{GeneralElectricGaussLaw}
  \mathrm{d}\,
  F'_2 
  =
  J_3
  \mathrlap{\,.}
\end{equation}
Here, the \emph{electric current density} $J_3$, which deforms the vacuum Gauss law \cref{FluxQuantizingHigherEMFields}, is spatially compactly supported. As such, it is defined on the spatial 1-point compactification $X^{1,3}_{\mathrm{cpt}}$, while the law \cref{GeneralElectricGaussLaw} holds on $\inlinetikzcd{X^{1,3} \ar[r, hook] \& X^{1,3}_{\mathrm{cpt}}}$. This way, \cref{GeneralElectricGaussLaw} does not imply that the total electric charge
\begin{equation}
  Q
  :=
  -e
  \int_{X^3_{\mathrm{cpt}}}
  J_3
\end{equation}
vanishes, and in fact we have that $-Q/e \in \mathbb{Z}$ is the net number of electrons.

But this means that the current density $J_3$ is classified by $K\bracket({\mathbb{Z}, 3})$. To trivialize on $X^{1,3}$ the corresponding classifying map equivalently means to lift through the \emph{path fibration} $\inlinetikzcd{P K\bracket({\mathbb{Z},3}) \ar[r, ->>] \& K\bracket({\mathbb{Z},3}) }$ like this:
\begin{equation}
  \label{RelativeElectricFluxQuantization}
  \begin{tikzcd}[row sep=15pt, column sep=large]
    X^{1,3}
    \ar[d, hook]
    \ar[r, dashed]
    &
    P K\bracket({\mathbb{Z},3})
    \ar[d, ->>]
    \\
    X^{1,3}_{\mathrm{cpt}}
    \ar[
      r,
      "{
        -Q/e
      }"
    ]
    &
    K\bracket({\mathbb{Z},3})
    \mathrlap{\,.}
  \end{tikzcd}
\end{equation}
This is hence the general relative form of electric flux quantization. Notice that the fiber of the path fibration is a $K\bracket({\mathbb{Z},2})$, so that \cref{RelativeElectricFluxQuantization} indeed reduces to \cref{ClassifyingOrdinaryCharge} where $Q = 0$.

\textbf{The general rule for relative flux quantization}, in generalization of \cref{GeneralRuleForFluxQuantization}, is essentially the following:
Consider an embedding $\inlinetikzcd{Y \ar[r, hook, "{ \Phi }"] \& X}$ with 
flux densities $F^{(i)}$ on $Y$
satisfying
\begin{equation}
  \label{GenericRelativeGaussLaw}
  \mathrm{d}\,
  F^{(i')}
  =
  P^{(i')}\bracket({
    F^{(1)}, F^{(2)} \cdots,
    \Phi^\ast J^{(1)}, \Phi^\ast J^{(2)}, \cdots
  })
  \mathrlap{\,,}
\end{equation}
for
\emph{background flux densities} $J^{(i)}$ 
on $X$ that themselves are subject to differential equations $\mathrm{d} J^{(i)} = P^{(i)}\bracket({J^{(1)}, \cdots, J^{(\sharp_J)}})$ as before for $F$ in \cref{GenericGaussLaw}. Then, given an admissible classifying space $\mathcal{B}$ for the $J$-s according to the corresponding version of \cref{GeneralRuleForFluxQuantization}, the admissible \emph{classifying fibrations} $\inlinetikzcd{ \mathcal{A} \ar[r,->>] \& \mathcal{B} }$ are those for which the real cohomology of their total spaces is computed by the cohomology of these differential relations when the $F^{(i')}$ and $J^{(i)}$ are regarded as abstract algebra generators $f^{(i')}$ and $j^{(i)}$, respectively:

\begin{equation}
  \label{GeneralRuleForRelativeFluxQuantization}
  \hspace{-3mm} 
  {\color{gray}\substack{
    \text{admissible}
    \\
    \text{relative}
    \\
    \text{global charge}
  }}
  \;\;
  \begin{tikzcd}[sep=15pt]
    Y
    \ar[
      d, hook, "{ \Phi }"
    ]
    \ar[r, dashed]
    &
    \mathcal{A}
    \ar[
      d,
      ->>,
      "{\,
        \mathcal{P}
      }"
    ]
    \\
    X
    \ar[r]
    &
    \mathcal{B}
  \end{tikzcd}
  \;\;\;
  \text{if}
  \;\;
  H^\bullet\bracket({
    \mathcal{A}
  })
  \overset{!}{\simeq}
  \frac
    {\mathrm{ker}(\mathrm{d})}
    {\mathrm{im}(\mathrm{d})}
  \,,
  \;
  \left\{
  \begin{aligned} 
  \mathrm{d}f^{(i')}
  & =
  P^{(i')}\bracket({
    f^{(1)}, \cdots,
    j^{(1)}, \cdots
  })
  \\
  \mathrm{d}j^{(i)}
  & =
  P^{(i)}\bracket({
    j^{(1)}, 
    j^{(2)}, 
    \cdots
  })
  \mathrlap{\,.}
  \end{aligned}
  \right.
\end{equation}

For example, in generalization of \cref{ChargeQuantizationInKTheory}, now  with $h_3$, $\mathrm{d} h_3 = 0$ denoting the generator of the cohomology of $K\bracket({\mathbb{Z},3})$ as in \cref{QuantizationOfHigherMagneticFlux}, consider the real cohomology of the classifying fibration
$\inlinetikzcd{\mathrm{KU}\sslash K\bracket({\mathbb{Z},2}) \ar[r] \& K\bracket({\mathbb{Z},3})}$ for  \emph{twisted K-theory}. This is 
 the $h_3$-twisted cohomology on even degree generators $f_{2\bullet}$, whence this qualifies as a classifying fibration for RR-flux $F_{2\bullet}$ twisted by $N$ units of NS-flux $H_3$ in type IIA supergravity:
\begin{equation}
  \label{FluxQuantizingTwistedRRFlux}
  {\color{gray}\substack{
    \text{electro/magnetic}
    \\
    \text{twisted}
    \\
    \text{Gauss laws}
  }}
  \;\;
  \begin{aligned}
    \mathrm{d}\, 
    F_{2\bullet + 2} 
    & = 
    H_3 \wedge F_{2\bullet}
    \\
    \mathrm{d} H_3
    & = 0    
    \,,
  \end{aligned}
  \;\;\;\;\;\;\;\;
  \begin{tikzcd}[sep=small]
    X^{1,9}
    \ar[r, dashed]
    \ar[d, hook]
    &
    \mathrm{KU}
    \sslash
    K\bracket({\mathbb{Z},2})
    \ar[d, ->>]
    \\
    X^{1,9}_{\mathrm{cpt}}
    \ar[r, "{ N }"]
    &
    K\bracket({\mathbb{Z},3})
  \end{tikzcd}
  \;\;
  {\color{gray}\substack{
    \text{relative}
    \\
    \text{global}
    \\
    \text{charge}
  }}
  \mathrlap{\,.}
\end{equation}
% While this Gauss law is no longer strictly linear it is actually still ``twisted linear'' and the quantizing cohomology theory, twisted K-theory, is still an abelian (albeit twisted) cohomology theory.

The example of interest in the following is this: To obtain the cohomology of the total space of the \emph{quaternionic Hopf fibration}  (cf. \cref{TheFibrations})
\begin{equation}
  \label{QuaternionicHopfFibration}
  \begin{tikzcd}[row sep=15pt]
    S^3
    \ar[r]
    &
    S^7
    \ar[d, ->>]
    \\
    & S^4
  \end{tikzcd}
\end{equation}
(which is concentrated in degree 7), starting with the generators $g_4$ and $g_7$ which present the cohomology of the 4-sphere  in \cref{FluxQuantizing11DSuGraEMField}, we need to add a generator $h_3$ which removes $g_4$ from cohomology. But this models just the Gauss law for the self-dual 3-form flux density on probe M5-brane worldvolumes $\inlinetikzcd{\Sigma^{1,5} \ar[r, hook] \& X^{1,10}}$, which thus has admissible flux quantization in relative Cohomotopy-twisted Cohomotopy, as shown on the right here (\cite[\S 7.3]{FSS20-H}\cite[(20)]{GSS25-M5}\cite[\S 3]{FSS21-Hopf}):
\begin{equation}
  \label{FluxQuantizingOnM5TheM3Flux}
  {\color{gray}\substack{
    \text{electro/magnetic}
    \\
    \text{twisted}
    \\
    \text{Gauss laws}
  }}
  \;\;
  \begin{aligned}
    \mathrm{d}\, 
    H_3 
    & = 
    \Phi^\ast G_4 
    \\
    \mathrm{d}\, G_7
    & = 
    \tfrac{1}{2}
    G_4 \wedge G_4
    \\
    \mathrm{d}\, G_4
    & = 
    0
    \,,
  \end{aligned}
  \;\;\;\;\;\;\;\;
  \begin{tikzcd}[
    column sep=small,
    row sep=17pt
  ]
    \Sigma^{1,5}
    \ar[r, dashed]
    \ar[d, hook, "{ \Phi }"]
    &
    S^7
    \ar[d, ->>]
    \\
    X^{1,10}
    \ar[r]
    &
    S^4
  \end{tikzcd}
  \;\;
  {\color{gray}\substack{
    \text{relative}
    \\
    \text{global}
    \\
    \text{charge}
  }}
  \mathrlap{\,.}
\end{equation}

\begin{figure}[htb]
\caption{\label{TheFibrations}
  The \emph{quaternionic Hopf fibration} $h_{\mathbb{H}}$ and 
  its factorization through the \emph{twistor fibration} $t_{\mathbb{C}}$ is given by sending $\mathbb{R}_+$-lines in quaternionic 2-space $\mathbb{H}^2$ first to the $\mathbb{C}$-lines which they span, and then further to the $\mathbb{H}$ lines which these span, using the canonical inclusions $\mathbb{R}_+ \subset \mathbb{C} \subset \mathbb{H}$ of the positive real numbers into the complex numbers and further into the quaternions (cf. \cite[\S 2]{FSS22-Twistorial}\cite[Rem. 5.2.16]{SS26-Orb}).
}
\centering
\adjustbox{rndfbox=4pt}{
\begin{tikzcd}[row sep=10pt]
  S^7
  \ar[
    r,
    ->>,
    "{ h_{\mathbb{C}} }"
  ]
  \ar[
    rr,
    uphordown,
    "{ h_{\mathbb{H}} }"
  ]
  \ar[
    d, equals
  ]
  & 
  \mathbb{C}P^3
  \ar[
    r,
    ->>,
    "{ t_{\mathbb{C}} }"
  ]
  \ar[
    d, equals
  ]
  &
  \overbrace{
    \mathbb{H}P^1
  }^{ S^4 }
  \ar[
    d, equals
  ]
  \\
  \bracket({
    \mathbb{H}^2 \! \setminus \! \{0\}
  })\big/\mathbb{R}_+
  \ar[
    r,
    ->>
  ]
  &
  \bracket({
    \mathbb{H}^2 \! \setminus \! \{0\}
  })\big/\mathbb{C}^\times
  \ar[
    r,
    ->>
  ]
  &
  \bracket({
    \mathbb{H}^2 \! \setminus \! \{0\}
  })\big/\mathbb{H}^\times
\end{tikzcd}
}
\end{figure}

There are various further variants of the situation \cref{FluxQuantizingOnM5TheM3Flux}:
For example, one may consider the \emph{double dimensional reduction} (via \emph{cyclification} of classifying spaces, \cite[\S 2.2]{BMSS2019}), of these flux-quantized M5s probing 11D SuGra, to D4-branes probing type IIA 10D SuGra (cf. \cite{Banerjee2026M5brane}). 
Another variant is to adjoin ``topological fluxes'' to the system; we consider this next.

%%%%%%%%%%%%%
\subsection
{Super Flux Quantization}
\label
{OnSuperFluxQuantization}
%%%%%%%%%%%%%%

Since the key to quantizing (higher) flux densities is -- as we have recalled --  their electro/magnetic Gauss laws, it is desirable to have a good grasp on how examples of these come about. 
A profound observation here is that, when seen on superspace, the Gauss laws imposed on super-flux densities are close to the full equations of motions.

For an 11D super-spacetime with supertorsion-free super-vielbein $(E,\Psi)$, consider the following super-flux lift of the ordinary flux densities $(G_4, G_7)$, where $(E,\Psi)$ is a super-coframe \cref{TheSuperembeddings} satisfying the torsion constraint \cref{11DSuGraTorsionConstraint}:
\begin{equation}
  \label{TheCFieldSuperFluxDensity}
  \begin{aligned}
    G_4^s
    & 
    :=
    \tfrac{1}{4!}
    (G_4)_{a_1 \cdots a_4}
    E^{a_1} \cdots E^{a_4}
    + 
    \tfrac{1}{2}
    \bracket({\,
      \overline{\Psi}
      \,\Gamma_{ab}\,
      \Psi
    })
    E^a E^b
    \\
    G_7^s
    & 
    :=
    \tfrac{1}{7!}
    (G_7)_{a_1 \cdots a_7}
    E^{a_1} \cdots E^{a_7}
    + 
    \tfrac{1}{5!}
    \bracket({\,
      \overline{\Psi}
      \,\Gamma_{a_1 \cdots a_5}\,
      \Psi
    })
    E^{a_1} \cdots E^{a_5}
    \mathrlap{\,.}
  \end{aligned}
\end{equation}
The form of the Gauss law \cref{FluxQuantizing11DSuGraEMField} imposed on \eqref{TheCFieldSuperFluxDensity}
is \emph{equivalent} to the equations of motion of 11D supergravity (\cite[Thm. 3.1]{GSS24-SuGra}):
\begin{equation}
  \label{OnShell11DSuGra}
  \left.
  \begin{aligned}
    \mathrm{d}\, 
    G_7^s 
    &=
    \tfrac{1}{2}
    G_4^s \wedge G_4^s
    \\
    \mathrm{d}\,
    G_4^s & = 0
  \end{aligned}
  \right\}
  \;\Leftrightarrow\;
  \text{on-shell 11D SuGra.}
\end{equation}

Similarly, with $(e, \psi)$ the induced super-coframe on the M5-brane worldvolume, \cref{DarbouxCoframeCondition}, the twisted Gauss law \cref{FluxQuantizingOnM5TheM3Flux} imposed on $H_3$ regarded on super-space,
\begin{equation}
  \label{TheSuperH3Flux}
  H_3^s 
  :=
  \tfrac{1}{3!}
  (H_3)_{a_1 a_2 a_3}
  e^{a_1} e^{a_2} e^{a_3}
  + 
  0
  \mathrlap{\,,}
\end{equation}
gives the (full non-linear) self-duality constraint and the equation of motion on the $H_3$ flux (\cite[Prop. 3.18]{GSS25-M5}):
\begin{equation}
  \label{EoMFor3Flux}
  \mathrm{d}\,
  H_3^s
  =
  \Phi^\ast
  G_4^s
  \Big\}
  \;\;\;
  \Leftrightarrow
  \;\;\;
  \text{on-shell self-dual tensor field.}
\end{equation}

While largely classical, this superspace perspective on supergravity offers an underappreciated, powerful approach to \emph{non-Lagrangian} theory building: 

\noindent {\it We may define theories by imposing Gauss laws on super-flux}, followed by flux quantization.

For example, consider adjoining, in this manner, to the $H_3^s$-flux on the M5 super-worldvolume a super 2-flux of the form
\begin{equation}
  \label{TheSuper2FormFlux}
  F_2^s
  :=
  \tfrac{1}{2}
  (F_2)_{a_1 a_2}
  e^{a_1} e^{a_2}
  \,.
\end{equation}
Now the non-degeneracy \cref{NonCriticalityCondition} of the torsion constraint on the M5-brane \cref{TorsionConstraintOnM5} implies (cf. \cite[p. 7]{SS25-Seifert}) that imposing on this super-flux the ordinary magnetic Gauss law (does not affect the previous equations of motion but otherwise) is equivalent to $F_2 = 0$, hence:
\begin{equation}
  \label{2FluxGaussLawIsAbelianChernSimons}
  \mathrm{d}\,
  F_2^s
  = 0
  \Big\}
  \;\;
  \Leftrightarrow
  \;\;
  \text{on-shell abelian Chern-Simons}
  \mathrlap{\,.}
\end{equation}
However, this has the remarkable consequence that we may add polynomials in a closed super-flux $F^s_2$ to the Gauss law
\eqref{EoMFor3Flux} for the self-dual tensor flux on superspace, \emph{without changing the local theory}, notably like this:
\begin{equation}
  \label{ModifiedSelfDualTensorField}
  \left.
  \begin{aligned}
    \mathrm{d}\,
    H_3^s
    & =
    \Phi^\ast
    G_4^s
    -
    \theta
    \,
    F^s_2 \wedge F^s_2
    \\
    \mathrm{d}\, 
    F^s_2
    & = 0
  \end{aligned}
  \right\}
  \;\;\;
  \Leftrightarrow
  \;\;\;
  \text{on-shell self-dual tensor field}
  ,
\end{equation}
for any \emph{theta angle} $\theta$.

This modified Gauss/Bianchi identity for $H_3$ systematically implies (\cite[p. 38]{SS25-Srni}\cite[(83)]{Banerjee2025-Potentials} following \cite[p. 22]{GSS24-SuGra}) that it is locally, on an open cover $\inlinetikzcd{\widehat{X} \ar[r,->>, "{\iota}"] \& X}$ of spacetime,  expressed in terms of \emph{gauge potential} forms $C_3$, $B_2$, $A_1$ on $\widehat X$ as
\begin{equation}
  \label{LocalStructureOfMTheory3Form}
  \iota^\ast H_3
  =
  C_3
  +
  \mathrm{d}B_2
  +
  \theta \, \mathrm{CS}(A_1)
  \mathrlap{\,,}
\end{equation}
(where $\mathrm{CS}(A_1) = A_1 \wedge \mathrm{d}A_1$ is the Chern-Simons form, here in the abelian case). Expressions of this form for the M-theory 3-form have previously been considered in 
\cite[(2.1)]{DonagiWijnholt2023}\cite[(2.1.15)]{FSS14-7D}\cite[\S 4.1]{FSS15-ModuliStack}, following \cite{DFM2007}, and earlier in \cite[(3.3)]{Evslin_2003}.

But while the local dynamics of flux densities does not change under this addition, when the Chern-Simons equations of motion hold \cref{2FluxGaussLawIsAbelianChernSimons}, the flux-quantized global theory does change substantially when $\theta \neq 0$, because the flux quantization changes, since $\inlinetikzcd{S^7 \ar[r,->>] \& S^4}$  \cref{FluxQuantizingOnM5TheM3Flux} is no longer an admissible classifying fibration.

To see how the flux quantization law gets modified for $\theta \neq 0$, notice that when imposed on abstract algebra generators $g_4$ and $f_2$, the law in \cref{ModifiedSelfDualTensorField} then says that $g_4$ represents the same cohomology class as $f_2^2$, whence the trivialization of $g_4^2$ in cohomology (by $g_7$) is then equivalent to the trivialization of $f_2^4$ in cohomology. But that is the relation on the cohomology of complex projective 3-space (cf. \cite[\S 5.3]{Menichi2015}), which is fibered over $S^4$ via the \emph{twistor fibration} (\cite[\S III.1]{Atiyah1978}, cf. \cite[p. 6]{FSS22-Twistorial}):
\begin{equation}
  \label{TwistorFibration}
  \begin{tikzcd}[row sep=15pt] 
    S^2
    \ar[r]
   &
    \mathbb{C}P^3 
      \ar[
        d, 
        ->>,
        "{ t_{\mathbb{C}} }"
      ] 
      \\
    &
    S^4.
  \end{tikzcd}
\end{equation}
Therefore this is, by  \cref{GeneralRuleForRelativeFluxQuantization}, an admissible relative flux quantization law for 11D SuGra with such ``magnetized'' M5-brane probes, for $\theta \neq 0$ (\cite[\S 4]{SS25-Seifert} following \cite{FSS22-Twistorial}, cf. \cite[\S 2]{SS25-Srni}):
\begin{equation}
  \label{FluxQuantizingOnMagnetizedM5TheM3Flux}
  {\color{gray}\substack{
    \text{twisted}
    \\
    \text{Gauss laws}
  }}
  \;\;
  \begin{aligned}
    \mathrm{d}\, 
    H_3 
    & = 
    \Phi^\ast G_4 
    -
    F_2 \wedge F_2
    \\
    \mathrm{d}\,
    F_2
    & = 0
    \\[4pt]
    \mathrm{d}\, G_7
    & = 
    \tfrac{1}{2}
    G_4 \wedge G_4
    \\
    \mathrm{d}\, G_4
    & = 
    0
    \,,
  \end{aligned}
  \;\;\;\;\;\;\;\;
  \begin{tikzcd}[
    column sep=small,
    row sep=30pt
  ]
    \Sigma^{1,5}
    \ar[r, dashed]
    \ar[d, hook, "{ \Phi }"]
    &
    \mathbb{C}P^3
    \ar[
      d, 
      ->>,
      "{ t_{\mathbb{C}} }"
    ]
    \\
    X^{1,10}
    \ar[r, dashed]
    &
    S^4
  \end{tikzcd}
  \;\;
  {\color{gray}\substack{
    \text{relative}
    \\
    \text{global}
    \\
    \text{charge}
  }}
  \mathrlap{\,.}
\end{equation}

\begin{added}
  To summarize:
  The field content on such a \emph{magnetized M5-branes} is locally (on a chart) the usual local field content, but globalized such that:
  \begin{enumerate}
  \item the local 2-form gauge potential $B_2$ of the self-dual tensor field  is globally accompanied by higher transition functions and coherence laws analogous to what is known for bundle gerbe connections, but here defining a class in twisted Cohomotopy instead of just a class in ordinary 3-cohomology;

  \item in addition there is a \emph{flat} $\mathrm{U}(1)$-connection field (the ``magnetization'') suitably twisted by the 2-form field. Since its flux density $F_2$ vanishes this does not show up in the usual local worldvolume theory, but globally it does contribute through topological effects (further discussed in \cref{OnASingularities}).
  \end{enumerate}
\end{added}

We next bring the M-string into this picture.

%%%%%%%%%%%%%%%%%%%%%%%%%%%%%%%%%%%%%%%
\section{M-String Flux Quantization}
\label
{OnMStringFluxQuantization}
 %%%%%%%%%%%%%%%%%%%%%%%%%%%%%%%%%%%%%%

%%%%%%%%%%%%%%%%%%%%%%%%
\subsection{The Law}
\label{TheFluxQuantizationOfMOnM5In11D}
%%%%%%%%%%%%%%%%%%%%%%%%

In the manner of \cref{TheSuper2FormFlux}, we may consider adjoining to the super-flux densities in the 11D bulk \cref{TheCFieldSuperFluxDensity} and on a M5-probe \cref{TheSuperH3Flux} also a super-flux on the M-string super-worldsheet \cref{TheSuperembeddings}, of the form
\begin{equation}
  \label{Super1FluxDensity}
  H_1^s
  :=
  (H_1)_a \StringVielbein^a
  + 
  0
\end{equation}
and subjected to the Gauss/Bianchi law
\begin{equation}
  \label{GaussLawForSuperH1}
  \mathrm{d}\,
  H^s_1
  =
  \phi^\ast F^s_2
  \mathrlap{\,.}
\end{equation}

Recall the subtle purely global/topological effect on the SuGra/brane dynamics that is controlled by such a further Gauss law: We have already seen in \cref{2FluxGaussLawIsAbelianChernSimons} that the Bianchi identity for $F_2^s$ forces it to vanish on-shell. Therefore, the presence of $F^s_2$ does not affect the on-shell dynamics at the level of flux densities, but it does crucially change the global/ topological behavior of the theory. Concretely, a flux density $F_2$ with $F_2 = 0$ induces a topological gauge field of abelian Chern-Simons type.

Analogously, the above Gauss law \cref{GaussLawForSuperH1} is, in turn, equivalent to the condition that $H_1$ vanishes, compatible with the superembedding result \cref{TildeH1Vanishes,H1}:
\begin{equation}
\label{H1VanishesByItsSuperBianchi}
\hspace{-.5cm} 
  \left.
  \begin{aligned}
    0 \;
    &
    \underset{\mathclap{\scalebox{.7}{\cref{GaussLawForSuperH1}}}}
      {=}
   \;
    \mathrm{d}\, 
    H^s_1
    -
    \phi^\ast F^s_2
    \\
    & 
    \underset{\mathclap{\scalebox{.7}{\cref{2FluxGaussLawIsAbelianChernSimons}}}}
      {=}
      \;
    \mathrm{d}\, 
    H^s_1
    \\
    & 
    \underset{\mathclap{\scalebox{.7}{\cref{Super1FluxDensity}}}}
      {=}
      \;
    \mathrm{d}
    \bracket({
      (H_1)_a \StringVielbein^a
    })
    \\
    & 
    \underset{\mathclap{\scalebox{.7}{\cref{TorsionConstraintOnString}}}}
      {=} 
    \phantom{+}
    \bracket({
      \nabla_b (H_1)_a
    })
    \StringVielbein^b \StringVielbein^a
    \\
    & 
    \phantom{=}\;
    +
    (H_1)_a
    \bracket({
      \phi^\ast 
      \CoefficientMatrix
    })^a_{a'}
    \bracket({
      \overline{\StringGravitino}
      \sigma^{a'}
      \StringGravitino
    })
  \end{aligned}
  \right\}
  \Leftrightarrow
  \left\{
  \begin{aligned}
    \bracket({
      \nabla_b (H_1)_a
    })
    \StringVielbein^b \StringVielbein^a
    & = 0
    \\
    (H_1)_a
    \bracket({
      \phi^\ast
      \CoefficientMatrix
    })^a_{a'}
    \bracket({
      \overline{\StringGravitino}
      \sigma^{a'}
      \StringGravitino
    })
    & = 0
  \end{aligned}
  \right\}
  \;
  \underset
    {\mathclap{\scalebox{.7}{\cref{NonCriticalityCondition}}}}
    {\Leftrightarrow}
    \;
  \Big\{
  H_1 = 0
  \mathrlap{\,.}
\end{equation}

In particular, adjoining the Gauss law \cref{GaussLawForSuperH1} 
again does not change the equations of motion of the underlying \{11D SuGra \!+\! branes\}-system, at the level of field strengths. 

But it does introduce topological effects. We may already see a hint of this at the level of the local gauge potentials, for notice that gauge potential 0-form $\lambda$ corresponding to $H_1$ in \eqref{2FluxGaussLawIsAbelianChernSimons} is characterized, on general grounds \cite[(97)]{Banerjee2025-Potentials} by the law
\begin{equation}
  \mathrm{d}\lambda
  = 
  H_1 - \phi^\ast(A_1)
  \mathrlap{\,,}
\end{equation}
where $A_1$ is the Chern-Simons field from \eqref{LocalStructureOfMTheory3Form}.
But for $H_1 = 0$ \eqref{H1VanishesByItsSuperBianchi} this becomes exactly the defining equation \cite[above (2.62)]{Wen1992}
\begin{equation}
  \label{AbelianWZWBianchi}
  \mathrm{d}\lambda = - \phi^\ast(A_1)
\end{equation}
for an abelian WZW field expected to emerge as a boundary effect of our abelian Chern-Simons field \eqref{2FluxGaussLawIsAbelianChernSimons} (we expand on this in \cite[\S 3.4]{SS26-BBC}, and we come back to this at the end, \cref{OnASingularities}).

\begin{added}
  In summary, the $H_1$ sector on the M-string appears as the boundary sector of the topological Chern-Simons $F_2$ sector in the bulk of the M5-brane. Both fluxes vanish on-shell, but as such induce topological gauge potential dynamics as seen in Chern-Simons theory with boundaries.
\end{added}

Indeed, adjoining $H_1$-flux as in \eqref{GaussLawForSuperH1} with \eqref{H1VanishesByItsSuperBianchi} does affect the admissible global topological completions of the field content on the sequence of brane embeddings, as follows.

Namely, with the above Gauss law \cref{Super1FluxDensity} adjoined to the system of Gauss laws \cref{FluxQuantizingOnMagnetizedM5TheM3Flux} on the magnetized M5 and in the bulk, the twistor fibration \cref{TwistorFibration} by itself is no longer an admissible relative flux quantization law, as in \cref{FluxQuantizingOnMagnetizedM5TheM3Flux}.
This is due to the general rule in \cref{GeneralRuleForRelativeFluxQuantization}: An admissible classifying fibration that accounts for \cref{Super1FluxDensity} needs its real cohomology to be computed on abstract algebra generators involving, besides the previous $f_2$, also a further generator $h_1$ whose differential is $\mathrm{d} h_1 = f_2$. This removes the class of $f_2$ from cohomology, contrary to the situation on $\mathbb{C}P^3$.

In fact, the only cohomology class that survives in this case is that of $g_7$, which may be understood as representing the unit cohomology class on the 7-sphere. Indeed, an admissible relative flux quantization of \cref{Super1FluxDensity} \emph{relative} to the twistorial situation \cref{FluxQuantizingOnMagnetizedM5TheM3Flux} is the sequence of fibrations which is obtained by factoring the quaternionic Hopf fibration \cref{QuaternionicHopfFibration} through the twistor fibration \cref{TwistorFibration} (cf. \cref{TheFibrations}) -- this follows by \cite[Lem. 2.13]{FSS22-Twistorial}:

\begin{equation}
  \label{FluxQuantizingOnMStringsInMagnetizedM5TheM3Flux}
  {\color{gray}\substack{
    \text{twisted}
    \\
    \text{Gauss laws}
  }}
  \;\;
  \begin{aligned}
    \mathrm{d}\,
    H_1
    & =
    \phi^\ast F_2    
    \\[4pt]
    \mathrm{d}\, 
    H_3 
    & = 
    \Phi^\ast G_4 
    -
    F_2 \wedge F_2
    \\
    \mathrm{d}\,
    F_2
    & = 0
    \\[4pt]
    \mathrm{d}\, G_7
    & = 
    \tfrac{1}{2}
    G_4 \wedge G_4
    \\
    \mathrm{d}\, G_4
    & = 
    0
    \,,
  \end{aligned}
  \;\;\;\;\;\;\;\;
  \begin{tikzcd}[
    column sep=small,
    row sep=30pt
  ]
    N^{1,1}
    \ar[r, dashed]
    \ar[
      d, 
      hook,
      "{ \phi }"
    ]
    &
    S^7
    \ar[
      d,
      ->>,
      "{
        h_{\mathbb{C}}
      }"
    ]
    \\[-15pt]
    \Sigma^{1,5}
    \ar[r, dashed]
    \ar[d, hook, "{ \Phi }"]
    &
    \mathbb{C}P^3
    \ar[
      d, 
      ->>,
      "{ t_{\mathbb{C}} }"
    ]
    \\
    X^{1,10}
    \ar[r, dashed]
    &
    S^4
  \end{tikzcd}
  \;\;
  {\color{gray}\substack{
    \text{relative}
    \\
    \text{global}
    \\
    \text{charge}
  }}
  \mathrlap{\,.}
\end{equation}

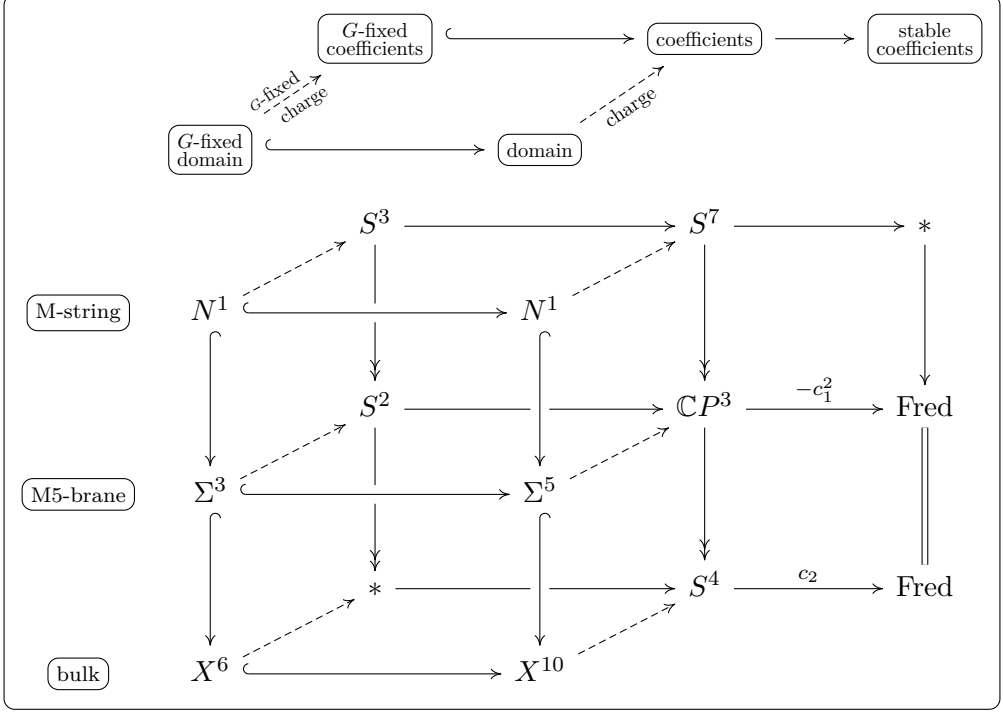
\begin{figure}[htb]
\caption{\label{OverviewDiagram}
  Overview{\protect\footnotemark}
  of the system of brane embeddings, of classifying fibrations and of charge classifying maps between these, for the M-brane on magnetized M5-branes in 11D SuGra, properly flux-quantized according to \cref{FluxQuantizingOnMStringsInMagnetizedM5TheM3Flux}.
  Solid maps are given (brane embeddings and classifying fibrations) while dashed maps are dynamical data (charge sectors of fields).
  \begin{added}
  In particular, the bottom square reflects cohomotopical C-field charge in the bulk, whose restriction to the orbi-singularity $X^6$ vanishes and (bottom right square) whose restriction to the magnetized M5-brane $\Sigma^5$ factors through the twistor fibration $t_{\mathbb{C}}$ (\cref{TheFibrations}) thereby exhibiting the self-dual worldvolume charge \cref{FluxQuantizingOnMagnetizedM5TheM3Flux}, whose further restriction to the M-string $N^1$ (top right square) factors through the Hopf fibration $h_{\mathbb{C}}$, thereby exhibiting the topological charge on the M-string \cref{FluxQuantizingOnMStringsInMagnetizedM5TheM3Flux}. The far right squares show the underlying charges in K-theory.
  \end{added}
}
\centering
\adjustbox{rndfbox=4pt}{
$
  \begin{tikzcd}[
    column sep=20pt,
    row sep=15pt
  ]
    &[-12pt]
    &
    \adjustbox{rndfbox=4pt}{$
    \substack{
      \scalebox{.7}{$G$-fixed}
      \\
      \scalebox{.7}{coefficients}
    }$}
    \ar[
      rr,
      hook
    ]
    &&
    \adjustbox{rndfbox=4pt}{$
    \substack{
      \scalebox{.7}{coefficients}
    }$}
    \ar[
      rr
    ]
    &[-10pt]&[-10pt]
    \adjustbox{rndfbox=4pt}{$
    \substack{
      \scalebox{.7}{stable}
      \\
      \scalebox{.7}{coefficients}
    }$}
    \\
    &
    \adjustbox{rndfbox=4pt}{$
    \substack{
      \scalebox{.7}{$G$-fixed}
      \\
      \scalebox{.7}{domain}
    }$}
    \ar[
      rr,
      hook
    ]
    \ar[
      ur,
      dashed,
      "{
        \scalebox{.65}{$G$-fixed}
      }"{pos=.3,sloped},
      "{
        \scalebox{.65}{charge}
      }"{pos=.55, sloped, swap}
    ]
    &&
    \adjustbox{rndfbox=4pt}{$
    \substack{
      \scalebox{.7}{domain}
    }$}
    \ar[
      ur,
      dashed,
      "{
        \scalebox{.7}{charge}
      }"{sloped, swap}
    ]
    \\[-10pt]
    &
    &
    S^3 
    \ar[rr]
    \ar[
      dd,
      ->>
    ]
    &&
    S^7
    \ar[rr]
    \ar[
      dd,
      ->>,
      "{
        h_{\mathbb{C}}
      }"{description}
    ]
    &&
    \ast
    \ar[dd]
    \\
    \adjustbox{rndfbox=4pt}{$
    \substack{
      \text{M-string}
    }$}
    &
    N^1
    \ar[
      rr,
      equals,
      crossing over
    ]
    \ar[
      dd,
      hook
    ]
    \ar[
      ur,
      dashed,
    ]
    &&
    N^1
    \ar[
      ur,
      dashed,
      "{
        \addedinline{
        \substack{
          \text{charge on}
        }
        }
      }"{sloped, pos=.4},
      "{
        \addedinline{
        \substack{
          \text{on M-string}
        }
        }
      }"{sloped, swap, pos=.4}
    ]
    \\
    &
    &
    S^2 
    \ar[rr]
    \ar[
      dd,
      ->>
    ]
    &&
    \mathbb{C}P^3
    \ar[
      dd,
      ->>,
      "{ 
        t_{\mathbb{C}} 
      }"{description}
    ]
    \ar[
      rr,
      "{ 
        - c_1^2 
      }"
    ]
    &&
    \mathrm{Fred}
    \ar[dd, equals]
    \\
    \adjustbox{rndfbox=4pt}{$
    \substack{
      \text{M5-brane}
    }$}
    &
    \Sigma^{3}
    \ar[
      rr,
      crossing over,
      hook
    ]
    \ar[
      dd,
      hook
    ]
    \ar[
      ur,
      dashed,
      "{
        \addedinline{
        \substack{
          \text{residual charge}
        }
        }
      }"{sloped},
      "{
        \addedinline{
        \substack{
          \;\;\text{@singularity}          
        }
        }
      }"{sloped, swap}
    ]
    &&
    \Sigma^{5}
    \ar[
      ur,
      dashed,
      "{
        \addedinline{
        \substack{
          \text{self-dual tensor}
          \\
          \text{charge on M5}
        }
        }
      }"{sloped, swap}
    ]
    \ar[
      from=uu,
      hook,
      crossing over
    ]
    \\
    &
    &
    \ast 
    \ar[rr]
    &&
    S^4
    \ar[
      rr, 
      "{ 
        c_2 
      }"
    ]
    &&
    \mathrm{Fred}
    \\
    \adjustbox{rndfbox=4pt}{$
    \substack{
      \text{bulk}
    }$}
    &
    X^{6}
    \ar[rr, hook]
    \ar[
      ur,
      dashed
    ]
    &
    & 
    X^{10}
    \ar[
      ur,
      dashed,
      "{
        \text{\addedinline{C-field charge}}
      }"{sloped, swap}
    ]
    \ar[
      from=uu, 
      hook,
      crossing over
    ]
  \end{tikzcd}
$
}
\end{figure}

\footnotetext{For further discussion of \cref{OverviewDiagram} see,
  for the right part:
  \cite[(97, 232)]{SS23-Mf},
  \cite[Fig. 6]{SS25-Orient}
  for the top left part: \cite[Fig. 5]{SS25-Orient},
  for the bottom rear part:
  \cite{SS25-Seifert},
  and for the bottom middle part:
  \cite[(25)]{SS25-Seifert},
  \cite[p. 13]{SS25-Srni}.
}

Let us highlight the consequences for the topological charges carried by these fields: 
Recall that for ordinary Dirac charge quantization \cref{ClassifyingOrdinaryCharge}, the total magnetic charge is given by the homotopy class of the classifying map from spacetime to $B \mathrm{U}(1)$:
\begin{equation}
  \text{magnetic charges}
  \in
  \pi_0\bracket({
    \mathrm{Map}\bracket({
      X^{1,3},
      B\mathrm{U}(1)
    })
  })
  \simeq
  H^2\bracket({
    X^{1,3};
    \mathbb{Z}
  })
  \mathrlap{\,.}
\end{equation}
Generally, for $\mathcal{A}$ the classifying space of a flux quantization law \cref{GeneralRuleForFluxQuantization}, we have:
\begin{equation}
  \text{charges}
  \in
  \mathrm{Map}\bracket({
    X, \mathcal{A}
  })
  \mathrlap{\,.}
\end{equation}

As we pass to relative flux quantization \cref{GeneralRuleForRelativeFluxQuantization}, the single classifying map is generalized to a pair of maps forming a commuting square with the domain embedding $\Phi$ and the classifying fibration $\mathcal{P}$. If we denote the topological space of such compatible pairs of classifying maps as%
\footnote{
  More precisely, the relative mapping space \cref{RelativeMappingSpace} is the \emph{fiber product} of $\mathrm{Map}\bracket({X,\mathcal{B}})$ with $\mathrm{Map}\bracket({Y,\mathcal{A}})$ with respect to their joint fibration over $\mathrm{Map}\bracket({Y,\mathcal{B}})$ (the former by precomposition with $\Phi$, the latter by postcomposition with $\mathcal{P}$). That this topological space has the correct homotopy type follows (by the existence of the Quillen model structure on topological spaces) from $\mathcal{P}$ being a (Serre-) fibration and $\Phi$ a cofibration. An analogous comment also holds for the doubly relative mapping space \cref{DoublyRelativeMappingSpace} below.
}
\begin{equation}
  \label{RelativeMappingSpace}
  \mathrm{Map}\bracket({
    \Phi, \mathcal{P}
  })
  :=
  \left\{
  \begin{tikzcd}[row sep=15pt]
    Y 
    \ar[r, dashed]
    \ar[d, hook, "{ \Phi }"]
    &
    \mathcal{A}
    \ar[
      d, 
      ->>, 
      "{ 
        \mathcal{P} 
      }"
    ]
    \\
    X 
    \ar[r, dashed]
    &
    \mathcal{B}
  \end{tikzcd}
  \,\right\},
\end{equation}
then the sets of relative charges of bulk/brane systems $\Phi$ whose fluxes are classified by the fibration $\mathcal{P}$ form the connected components of this relative mapping space:
\begin{equation}
  \text{relative charges}
  \in
  \pi_0\, 
  \mathrm{Map}\bracket({
    \Phi, 
    \mathcal{P}
  })
  \mathrlap{\,.}
\end{equation}

In this manner, if we next have a sequence $\inlinetikzcd{ {} \ar[r, hook, shorten=-2pt, "{\phi}"] \& {} \ar[r, hook, shorten=-2pt, "{\Phi}"] \& {} }$ of brane embeddings, as in \cref{TheSuperembeddings}, 
and a sequence $\inlinetikzcd{ {} \ar[r, ->>, shorten=-2pt, "{\wp}"] \& {} \ar[r, ->>, shorten=-2pt, "{ \mathcal{P} }"] \& {}  }$ of classifying fibrations for their relative flux quantization, as in \cref{FluxQuantizingOnMStringsInMagnetizedM5TheM3Flux}, then the connected components of the ``doubly relative mapping space''
\begin{equation}
  \label{DoublyRelativeMappingSpace}
  \mathrm{Map}\bracket({
    \bracket({\phi,\Phi})
    ,
    \bracket({\wp,\mathcal{P}})
  })
  :=
  \left\{
  \begin{tikzcd}[sep=small]
    \phantom{A}
    \ar[
      r,
      shorten=-4pt,
      dashed
    ]
    \ar[
      d,
      hook,
      shorten=-4pt,
      "{ \phi }"
    ]
    &
    \phantom{A}
    \ar[
      d,
      ->>,
      shorten=-4pt,
      "{ \wp }"
    ]
    \\
    \phantom{A}
    \ar[
      d,
      ->>,
      shorten=-4pt,
      "{ \Phi }"
    ]
    \ar[
      r, 
      shorten=-4pt,
      dashed
    ]
    & 
    \phantom{A}
    \ar[
      d,
      ->>,
      shorten=-4pt,
      "{ \mathcal{P} }"
    ]
    \\
    \phantom{A}
    \ar[
      r,
      shorten=-4pt,
      dashed
    ]
    &
    \phantom{A}
  \end{tikzcd}
  \,\right\}
\end{equation}
are the doubly relative charges that may be carried by this system:
\begin{equation}
  \label{DoublyRelativeCharges}
  \text{doubly relative charges}
  \in
  \pi_0\, 
  \mathrm{Map}\bracket({
    \bracket({\phi,\Phi})
    ,
    \bracket({\wp,\mathcal{P}})
  })
  \mathrlap{\,.}
\end{equation}
We highlight that all this is the topological incarnation of iterated relative Gauss/Bianchi identities \cref{GenericRelativeGaussLaw}:

\begin{added}
  The commutativity of the squares of classifying maps in \cref{DoublyRelativeMappingSpace} is the topological incarnation of the appearance of new fluxes on brane embeddings $\phi$, $\Phi$
  whose classifying spaces for trivial bulk charges would be the fibers of the classifying fibrations $\wp$, $\mathcal{P}$, but 
  whose Bianchi identities are actually twisted by these bulk fluxes. 

  For example, when the bulk C-field charge happens to vanish, in that its classifying map factors through the point, then the bottom square in \cref{FluxQuantizingOnMStringsInMagnetizedM5TheM3Flux} factors through the $S^2$-fiber of the twistor fibration, exhibiting the self-dual 3-flux $H_3$ as quantized as a Hopfion charge:

\begin{equation}
    \begin{tikzcd}[
      ampersand replacement=\&
    ]
      \Sigma^5
      \ar[
        rr,
        dashed,
        uphordown,
        "{
          \text{M5 worldvolume charge}
        }"{swap}
      ]
      \ar[
        r,
        dashed
      ]
      \ar[
        d,
        hook,
        "{ \Phi }"
      ]
      \&
      S^2
      \ar[r]
      \ar[d]
      \ar[
        dr,
        phantom,
        "{ \lrcorner }"{pos=.1}
      ]
      \&
      \mathbb{C}P^3
      \ar[
        d,
        ->>,
        "{ t_{\mathbb{C}} }"
      ]
      \\
      X^{1,10}
      \ar[r]
      \ar[
        rr,
        downhorup,
        "{
          \text{vanishing C-field charge}
        }"
      ]
      \&
      \ast
      \ar[r]
      \&
      S^4
      \mathrlap{\,.}
    \end{tikzcd}
\end{equation}

  The commuting squares in \cref{DoublyRelativeMappingSpace} are the twisted generalization of this situation to the case that the (iterated) background charge does not vanish, reflecting the worldvolume Bianchi identities twisted by background fluxes.

  The rational homotopy theory providing these relations between relative twisted Bianchi identities and their flux quantization by such commuting squares of classifying maps is established in \cite{FSS22-Twistorial}, reviewed in \cite[\S12]{FSS23-Char}.

\end{added}

%%%%%%%%%%%%%%%%%%%%%%%%%%
\subsection{Orbifolding}
\label{OnOrbifolding}
%%%%%%%%%%%%%%%%%%%%%%%%%%

To see the implications of this doubly relative flux quantization on M-branes \cref{FluxQuantizingOnMStringsInMagnetizedM5TheM3Flux}, it is expedient to first consider the situation on A-type orbifold singularities, where the setup simplifies and can be related to more familiar topological effects (cf. \S\ref{OnASingularities}).

The \emph{orbifold cohomology} (cf. \cite{SS26-Orb}) of global orbifold quotients that we are concerned with here is modeled by equipping both the domain spaces and the classifying spaces with continuous actions of a group $G$ and then constraining the classifying maps $c$ to be \emph{$G$-equivariant}
\begin{equation}
  \text{$G$-orbifold charges}
  \;\in\;
  \pi_0\,
  \mathrm{Map}\bracket({X,\mathcal{A}})^G
  :=
  \pi_0
  \Bigg\{\!
  \adjustbox{raise=-6pt}{
  \begin{tikzcd}
    X
    \ar[
      out=60,
      in=180-60,
      shift right=-1pt,
      looseness=4,
      "{ 
        \,\mathclap{G}\,
      }"{description}
    ]
    \ar[
      r,
      dashed,
      "{ c }"
    ]
    &
    \mathcal{A}
    \ar[
      out=60,
      in=180-60,
      looseness=4,
      shift right=-1pt,
      "{ 
        \,\mathclap{G}\,
      }"{description}
    ]
  \end{tikzcd}
  }
  \!\!\Bigg\}
  \mathrlap{,}
\end{equation}
in that for all $x \in X$ and $g \in G$ we have:
\begin{equation}
  c(g \cdot x) = g \cdot c(x)
  \mathrlap{\,.}
\end{equation}

An important consequence of this equivariance condition is that \emph{$G$-fixed points} 
\begin{equation}
  \label{GFixedLocus}
  X^G
  :=
  \bracketmid
    \{{x \in X}{\forall_g : g \cdot x = x}\}
  \subset 
  X
\end{equation}
must be mapped to $G$-fixed points:
\begin{equation}
  \label{InducedMapOnFixedPoints}
  \begin{tikzcd}[row sep=small]
    X
    \ar[
      out=60,
      in=180-60,
      shift right=-1pt,
      looseness=4,
      "{ 
        \,\mathclap{G}\,
      }"{description}
    ]
    \ar[
      r,
      dashed,
      "{ c }"
    ]
    &
    \mathcal{A}
    \ar[
      out=60,
      in=180-60,
      looseness=4,
      shift right=-1pt,
      "{ 
        \,\mathclap{G}\,
      }"{description}
    ]
    \\
    X^G
    \ar[u, hook]
    \ar[
      r,
      dashed,
    ]
    &
    \mathcal{A}^G
    \mathrlap{\,.}
    \ar[u, hook]
  \end{tikzcd}
\end{equation}
In particular, the (iterated) mapping spaces \eqref{DoublyRelativeMappingSpace} between $G$-spaces become themselves $G$-spaces by the evident $G$-conjugation action, and their $G$-fixed loci are the (iterated) spaces of equivariant maps:
\begin{equation}
  \label{DoublyRelativeEquivariantMappingSpace}
  \mathrm{Map}\bracket({
    \bracket({\phi,\Phi})
    ,
    \bracket({\wp,\mathcal{P}})
  })^G
  :=
  \left\{
  \begin{tikzcd}[sep=17pt]
    \phantom{A}
    \ar[
      out=53+45,
      in=180-53+45,
      shift left=6pt,
      looseness=4,
      "{ 
        \,\mathclap{G}\;
      }"{description}
    ]
    \ar[
      r,
      shorten=-4pt,
      dashed
    ]
    \ar[
      d,
      hook,
      shorten=-4pt,
      "{ \phi }"
    ]
    &
    \phantom{A}
    \ar[
      out=53-45,
      in=180-53-45,
      shift left=6pt,
      looseness=4,
      "{ 
        \,\mathclap{G}\,
      }"{description}
    ]
    \ar[
      d,
      ->>,
      shorten=-4pt,
      "{ \wp }"
    ]
    \\
    \ar[
      out=60+90,
      in=180-60+90,
      shift left=5pt,
      looseness=3.4,
      "{ 
        \,\mathclap{G}\;
      }"{description}
    ]
    \phantom{A}
    \ar[
      d,
      ->>,
      shorten=-4pt,
      "{ \Phi }"
    ]
    \ar[
      r, 
      shorten=-4pt,
      dashed
    ]
    & 
    \phantom{A}
    \ar[
      out=60-90,
      in=180-60-90,
      shift left=5pt,
      looseness=3.4,
      "{ 
        \,\mathclap{G}\;
      }"{description}
    ]
    \ar[
      d,
      ->>,
      shorten=-4pt,
      "{ \mathcal{P} }"
    ]
    \\
    \phantom{A}
    \ar[
      out=53+135,
      in=180-53+135,
      shift left=5pt,
      looseness=4,
      "{ 
        \,\mathclap{G}\;
      }"{description}
    ]
    \ar[
      r,
      shorten=-4pt,
      dashed
    ]
    &
    \phantom{A}
    \ar[
      out=53-135,
      in=180-53-135,
      shift left=5pt,
      looseness=4,
      "{ 
        \,\mathclap{G}\;
      }"{description}
    ]
  \end{tikzcd}
  \,\right\}
\end{equation}
The joint homotopy classes of such equivariant maps are the (relative) orbifold charges (cf. \cite[(3)]{SS20-Tad}\cite[Fig. 7]{SS25-Orient}\cite{SS26-Orb}), in generalization of \eqref{DoublyRelativeCharges}:
\begin{equation}
  \label{DoublyEquivariantOrbifoldCharges}
  \text{Doubly relative orbifold charges}
  \in
  \pi_0 
  \,
  \mathrm{Map}\bracket({
    \bracket({\phi,\Phi})
    ,
    \bracket({\wp,\mathcal{P}})
  })^G.
\end{equation}

%%%%%%%%%%%%%%%
\subsection
{A-Singularities}
\label
{OnASingularities}
%%%%%%%%%%%%%%

Specifically for $A_{n-1}$-type orbi-singularities, the equivariance group is finite cyclic
$$
  G := \CyclicGroup{n}
  \mathrlap{\,,}
$$ 
and the action on the domain space is locally a product with $\mathbb{C}^2$ on which $[k] \in \CyclicGroup{n}$ acts as follows:
\begin{equation}
  \label{TheATypeAction}
  \begin{tikzcd}[
    column sep=-3pt
  ]
    \mathbb{H}
    \ar[
      out=60,
      in=180-60,
      shift left=1pt,
      looseness=4,
      "{ 
        \,\mathclap{G}\,
      }"{description}
    ]
    &\simeq_{{}_\mathbb{R}}&
    \mathbb{C}
    \ar[
      out=60,
      in=180-60,
      shift left=1pt,
      looseness=4,
      "{ 
        \,\mathclap{G}\,
      }"{description}
    ]
    &\times&
    \overline{\mathbb{C}}
    \ar[
      out=62,
      in=180-62,
      shift left=1pt,
      looseness=3,
      "{ 
        \,\mathclap{G}\,
      }"{description}
    ]
  \end{tikzcd}
  \hspace{15pt}
  :
  \hspace{15pt}
  \begin{tikzcd}
    {[k]} \cdot (z_1, z_2)
    :=
    \bracket({
      e^{2\pi\mathrm{i} k/n}
      z_1,
      \,
      e^{-2\pi\mathrm{i} k/n}
      z_2
    })
    \mathrlap{\,.}
  \end{tikzcd}
\end{equation}
As indicated, this action is in fact right $\mathbb{H}$-linear with respect to the identification $\mathbb{C}^2 \simeq_{{}_{\mathbb{R}}} \!\mathbb{H}$, being the left multiplication action with unit quaternions in the image of the inclusion
\begin{equation}
  \mathrm{U}(1)
  \subset 
  \mathrm{SU}(2)
  \simeq
  S({\mathbb{H}})
  \mathrlap{\,.}
\end{equation}
Therefore we may take the $G$-action on the classifying fibrations \cref{FluxQuantizingOnMStringsInMagnetizedM5TheM3Flux} to be ``of the same form'' as on spacetime, namely with $G$ acting as in \cref{TheATypeAction} on one of the two $\mathbb{C}^2 \simeq \mathbb{H}$-factors:
\begin{equation}
  \begin{tikzcd}[
    column sep=-5pt,
    row sep=6pt
  ]
    &&
    S^7
    \ar[rrrr, ->>, "{ t_{\mathbb{C}} }"]
    \ar[d, equals]
    \ar[
      out=60,
      in=180-60,
      shift left=1pt,
      looseness=4,
      "{ 
        \,\mathclap{G}\,
      }"{description}
    ]
    &&[27pt]&&
    \mathbb{C}P^3
    \ar[
      out=60,
      in=180-60,
      shift left=1pt,
      looseness=4,
      "{ 
        \,\mathclap{G}\,
      }"{description}
    ]
    \ar[rrrr, ->>, "{ t_{\mathbb{H}} }"]
    \ar[d, equals]
    &&[27pt]&&
    S^4
    \ar[
      out=60,
      in=180-60,
      shift left=1pt,
      looseness=4,
      "{ 
        \,\mathclap{G}\,
      }"{description}
    ]
    \ar[d, equals]
    \\
    \big(
    \mathbb{C}^2 
    \ar[
      out=60-180,
      in=-60,
      shift left=1pt,
      looseness=4,
      "{ 
        \,\mathclap{G}\,
      }"{description}
    ]
    \!&\times&
    \mathbb{C}^2
    &\!\setminus \{0\}
    \big)\big/\mathbb{R}_+
    \ar[r,->>]
    &
    \big(
    \mathbb{C}^2 
    \ar[
      out=60-180,
      in=-60,
      shift left=1pt,
      looseness=4,
      "{ 
        \,\mathclap{G}\,
      }"{description}
    ]
    &\times& \!\!
    \mathbb{C}^2
    &\!\!\!\!\setminus \{0\}
    \big)\big/\mathbb{C}^\times
    \ar[r, ->>]
    &
    \big(
    \mathbb{C}^2 
    \ar[
      out=60-180,
      in=-60,
      shift left=1pt,
      looseness=4,
      "{ 
        \,\mathclap{G}\,
      }"{description}
    ]
    &\times&
    \mathbb{C}^2
    &\!\setminus \{0\}
    \big)\big/\mathbb{H}^\times
    \mathrlap{\,.}
  \end{tikzcd}
\end{equation}
It is manifest from this definition that the corresponding fibration of fixed loci \cref{InducedMapOnFixedPoints} is the map from $S^3$ to the point, factored through the ordinary complex Hopf fibration $h_{\mathbb{C}}$:
\begin{equation}
  \label{FixedLocusInClassifyingFibrations}
  \begin{tikzcd}[
    column sep=35pt,
    row sep=10pt
  ]
    (S^7)^G
    \ar[
      r,
      ->>,
      "{ h_{\mathbb{C}}^G }"
    ]
    \ar[
      d,
      equals
    ]
    &
    (\mathbb{C}P^3)^G
    \ar[
      d,
      equals
    ]
    \ar[
      r,
      ->>,
      "{
        t_{\mathbb{C}}^G
      }"
    ]
    &
    (S^4)^G
    \ar[
      d,
      equals
    ]
    \\
    S^3 
    \ar[
      d,
      equals
    ]
    \ar[
      r,
      ->>,
      "{ h_{\mathbb{C}} }"
    ]
    &
    S^2
    \ar[
      d,
      equals
    ]
    \ar[
      r,
      ->>
    ]
    &
    \ast
    \ar[
      d,
      equals
    ]
    \\
    \bracket({
      \mathbb{C}^2
      \setminus
      \{0\}
    })\big/\mathbb{R}_+
    \ar[r, ->>]
    &
    \bracket({
      \mathbb{C}^2
      \setminus
      \{0\}
    })\big/\mathbb{C}^\times
    \ar[
      r,
      ->>
    ]
    &
    \bracket({
      \mathbb{C}^2
      \setminus
      \{0\}
    })\big/\mathbb{H}^\times
    \mathrlap{\,.}
  \end{tikzcd}
\end{equation}

Consider then (as in \cite[p. 5]{CouzensLuscherSparks2025}) an M5-brane worldvolume wrapping the half-space 
$$
  \begin{tikzcd}[
    column sep=-2pt
  ]
  \mathbb{C} 
    \ar[
      out=60,
      in=180-60,
      shift right=-1pt,
      looseness=4,
      "{ 
        \,\mathclap{G}\,
      }"{description}
    ]
    &\subset& 
  \mathbb{C}^2
    \ar[
      out=60,
      in=180-60,
      shift right=-1pt,
      looseness=3,
      "{ 
        \,\mathclap{G}\,
      }"{description}
    ]
  \end{tikzcd}
$$
of the $A$-type spacetime singularity \cref{TheATypeAction}, with the M-string embedded transverse to the singularity:
$$
  \begin{tikzcd}[column sep=-4pt]
    N^{1,1}
    \ar[
      d, 
      hook,
      "{ \phi }"
    ]
    &\defneq&
    N^{1,1}
    \ar[
      d, 
      hook,
    ]
    \\
    \Sigma^{1,5}
    \ar[
      d,
      hook,
      "{ \Phi }"
    ]
    &:=&
    \Sigma^{1,3}
    \ar[
      d,
      equals
    ]
    &
    \times
    &
    \mathbb{C}
    \ar[
      out=60,
      in=180-60,
      shift right=-1pt,
      looseness=4,
      "{ 
        \,\mathclap{G}\,
      }"{description}
    ]
    \ar[
      d,
      equals
    ]
    \\
    X^{1,10}
    &:=&
    \Sigma^{1,3}
    &
    \times
    &
    \mathbb{C}
    \ar[
      out=60-180,
      in=-60,
      shift right=-1pt,
      looseness=4.7,
      "{ 
        \,\mathclap{G}\,
      }"{description}
    ]
    &\times&
    \overline{\mathbb{C}}
    \ar[
      out=60-180,
      in=-60,
      shift right=-1pt,
      looseness=4.2,
      "{ 
        \,\mathclap{G}\,
      }"{description}
    ]
    &\times&
    X^3.
  \end{tikzcd}
$$

This configuration is evidently $G$-equivariantly homotopic to its $G$-fixed locus \eqref{GFixedLocus} (by linearly contracting the non-compact singularity to the point, $\mathbb{C}^2 \sim \ast$). Since the (iterated) equivariant mapping spaces \eqref{DoublyRelativeEquivariantMappingSpace} are homotopy-invariant under equivariant homotopy equivalence, the orbifold charges \eqref{DoublyEquivariantOrbifoldCharges} reduce, by \eqref{InducedMapOnFixedPoints}, to plain charges \eqref{DoublyRelativeCharges} on the fixed locus with $G$-fixed coefficients \eqref{FixedLocusInClassifyingFibrations}, cf. \cite[p. 7]{SS25-Seifert}\cite[\S 4]{SS25-TEC}:
\begin{equation}
  \left\{
  \hspace{-10pt}
  \vrule width 0pt depth 50pt height 50pt
  \smash{
  \begin{tikzcd}[
    ampersand replacement=\&,
    sep=20pt
  ]
    N^{1,1}
    \ar[
      out=53+45,
      in=180-53+45,
      shift left=3pt,
      looseness=4,
      "{ 
        \,\mathclap{G}\;
      }"{description}
    ]
    \ar[
      r,
      dashed
    ]
    \ar[
      d,
      hook,
      "{ \phi }"
    ]
    \&
    S^7
    \ar[
      out=53-45,
      in=180-53-45,
      shift left=6pt,
      looseness=4,
      "{ 
        \,\mathclap{G}\,
      }"{description}
    ]
    \ar[
      d,
      ->>,
      "{ h_{\mathbb{C}} }"
    ]
    \\
    \Sigma^{1,5}
    \ar[
      out=60+90,
      in=180-60+90,
      shift left=5pt,
      looseness=3.4,
      "{ 
        \,\mathclap{G}\;
      }"{description}
    ]
    \ar[
      d,
      ->>,
      "{ \Phi }"
    ]
    \ar[
      r, 
      dashed
    ]
    \& 
    \mathbb{C}P^3
    \ar[
      out=60-90,
      in=180-60-90,
      shift left=5pt,
      looseness=3.4,
      "{ 
        \,\mathclap{G}\;
      }"{description}
    ]
    \ar[
      d,
      ->>,
      "{ t_{\mathbb{C}} }"
    ]
    \\
    X^{1,10}
    \ar[
      out=53+135,
      in=180-53+135,
      shift left=5pt,
      looseness=4,
      "{ 
        \,\mathclap{G}\;
      }"{description}
    ]
    \ar[
      r,
      dashed
    ]
    \&
    S^4
    \ar[
      out=53-135,
      in=180-53-135,
      shift left=2pt,
      looseness=4,
      "{ 
        \,\mathclap{G}\;
      }"{description}
    ]
  \end{tikzcd}
  }
  \hspace{-4pt}
  \,\right\}
  \simeq
  \left\{
  \begin{tikzcd}[
    sep=15pt
  ]
    N^{1,1}
    \ar[r, dashed]
    \ar[
      d,
      "{ \phi }"
    ]
    &
    S^3
    \ar[
      d,
      "{ h_{\mathbb{C}} \,}"
    ]
    \\
    \Sigma^{1,3}
    \ar[r, dashed]
    &
    S^2
  \end{tikzcd}
  \right\}
  \mathrlap{.}
\end{equation}

But this is, on the right, exactly the \emph{Hopfion}-like anyonic charge structure of Fractional Quantum (Anomalous) Hall (FQ(A)H) systems (by \cite{SS25-FQH,SS25-FQAH,SS25-ISQS29}), with the M-string now playing the role of a gapped nodal line \cite[\S 2.2, Fig. 2.3]{SS26-Orb} on which \eqref{AbelianWZWBianchi} exhibits edge mode excitations \cite[\S 2.5]{Wen1992}\cite[\S 6.1.2]{Tong2016}; we further discuss this in \cite{SS26-BBC}. This phenomenon, that anyonic topological order may be \emph{geometrically engineered} (cf. \cite{Duplij2017}) on M5-branes wrapping (here: orbisingular) Seifert fibrations, was also argued (by very different, informal Lagrangian means) in \cite{ChoGangKim2020}.

%%%%%%%%%%%%%%%%%%%%%%%%%%%%%%%%%%%
\section{Conclusion}
\label{Conclusion}
%%%%%%%%%%%%%%%%%%%%%%%%%%%%%%%%%%%

Discussions of field theories, notably in the context of supergravity with brane probes, traditionally tend to restrict attention to the local field content given by flux densities defined on a single chart of spacetime, ignoring the global topological nature of the (higher) gauge fields. We have recalled in \S\ref{OnProperFluxQuantization} how global completion requires a choice of \emph{flux quantization} given by (fibrations of) topological \emph{classifying spaces} for the charges. This involves a great freedom of choice, reflecting a rich and largely unexplored space of flux-quantized Gauss laws/Bianchi identities. Ordinary electromagnetic Dirac charge quantization as well as K-theory quantization of RR-flux are just two examples of a general theory.

With a general theory of flux quantization in hand, we may in particular treat more subtle situations such as \emph{relative flux quantization} where bulk fluxes are probed by branes that carry their own worldvolume flux, a crucial step that has traditionally not found attention. The key example here is the M5-brane with its self-dual worldvolume flux probing the bulk of 11D supergravity with its C-field flux.

In this vein, here we went one step further and considered the situation of sequences of iterated brane species, concretely the case where the M5-brane probing the 11D bulk is itself probed by an M-string (M1), for which we developed relevant aspects of its superembedding construction in \S\ref{OnMStringSuperembedding}. 

This development is logically independent from the web of conjectures in the string theory folklore: We are using the on-shell superspace formulation of supergravity with super $p$-branes, and consider its systematic global completion by flux quantization; all our statements stand rigorously on their own feet.

Concretely, in \S\ref{OnMStringFluxQuantization} we have explored one possible flux quantization law, \eqref{FluxQuantizingOnMStringsInMagnetizedM5TheM3Flux}, for M1-probes of M5-probes in 11D,
highlighting its resemblance to previous conjectures about the structure of M-theory 3-forms \cref{LocalStructureOfMTheory3Form} and its implication of previously conjectured topological order engineered on M5-branes wrapping Seifert fibers (\S\ref{OnASingularities}) --- now enhanced by the presence of gapped nodal lines represented by the M-string.

This choice crucially involves another underappreciated aspect of global field completion: Since the non-Lagrangian on-shell superspace construction of 11D SuGra and its branes prescribes only flux densities and not their gauge potentials (which instead are brought into existence by compatible flux quantization), there is actually the freedom to adjoin Chern-Simons-type gauge fields, whose flux densities identically vanish on-shell \eqref{2FluxGaussLawIsAbelianChernSimons}\eqref{H1VanishesByItsSuperBianchi}. These are invisible to the local physics but their topological effect shows up in the global completion of the theory. This phenomenon is implicit in previous proposals, cf. \eqref{LocalStructureOfMTheory3Form}, but remains underappreciated. 

We have used this freedom in global completion to exhibit, in \S\ref{TheFluxQuantizationOfMOnM5In11D}, a Chern-Simons-like 1-form flux $H_1$ on the M-string, which is not visible from the usual local superembedding construction but is consistent with it. This choice is motivated by and appears rather suggestive due to the fact (\S\ref{OnASingularities}) that it implies on A-type orbisingularities, where the M5-brane engineers FQAH-type topological order, that the M-string engineers the gapped nodal lines of the system. 

However, we emphasize that this is just one choice out of an infinite space of other admissible choices. For instance, if we omitted the (on-shell vanishing) source term on the right of \eqref{GaussLawForSuperH1} then instead of the classifying space $S^7$ in \eqref{FluxQuantizingOnMStringsInMagnetizedM5TheM3Flux}, non-trivially Hopf-fibered over $\mathbb{C}P^3$, an admissible classifying space at that stage would instead be the trivial circle bundle space $\mathbb{C}P^3 \times S^1$, with less interesting topological consequences.

Hence, independent of the concrete choice of flux quantization on the M-string explored here, our discussion highlights that \emph{some} such choice is crucially to be made when passing from the traditional local discussion of fluxes on M-branes to their globally completed theories. Exploring this realm of globally completed brane-probed supergravity remains a wide-open field, to which we suggest we have opened the door a little.

%%%%%%%%%%%%%%%%%%%%%%%%%
\acknowledgments{
  HS and US acknowledge support by {\it Tamkeen UAE} under the {\it Abu Dhabi Research Institute Grant} {\tt CG008}.
}

%%%%%%%%%%%%%%%%%%%%%%%%%%%%%%
\bibliographystyle{JHEP}
\bibliography{refs-abbreviated}
%%%%%%%%%%%%%%%%%%%%%%%%%%%%%

%%%%%%%%%%%%%%%%%%
\end{document}